\documentclass[aps,twocolumn,showpacs,preprintnumbers,floats]{revtex4}\def\@cite#1#2{\textsuperscript{[{#1\if@tempswa , #2\fi}]}}
\usepackage{mathrsfs}

\usepackage{longtable,lscape}
\usepackage{savesym}
\usepackage{txfonts}
\savesymbol{iint}
\savesymbol{iiint}
\savesymbol{iiiint}
\savesymbol{idotsint}
\usepackage{amssymb}
\usepackage{amsmath}
\restoresymbol{TXF}{iint}
\restoresymbol{TXF}{iiint}
\restoresymbol{TXF}{iiiint}
\restoresymbol{TXF}{idotsint}
\usepackage{indentfirst}
\usepackage{graphicx,booktabs}
\usepackage{braket}
\usepackage{multirow}
\usepackage{color}
\usepackage{subfigure}
\usepackage{float}
\usepackage{bm}
\usepackage{epsfig}
\usepackage[colorlinks, citecolor=blue,anchorcolor=red,menucolor=red, linkcolor=red,filecolor=red,urlcolor=blue,frenchlinks=red]{hyperref}

\begin{document}
	
	\title{Possible explanations of the observed $\Lambda_c$ resonances}
	\author{Yu-Bin Zhang$^{1}$, Li-Ye Xiao$^{1}$~\footnote {E-mail: lyxiao@ustb.edu.cn}, Xian-Hui Zhong$^{2,3}$~\footnote {E-mail: zhongxh@hunnu.edu.cn}}
\affiliation{ 1)Institute of Theoretical Physics, University of Science and Technology Beijing,
Beijing 100083, China}
\affiliation{ 2) Department of
Physics, Hunan Normal University, and Key Laboratory of
Low-Dimensional Quantum Structures and Quantum Control of Ministry
of Education, Changsha 410081, China }
\affiliation{ 3) Synergetic
Innovation Center for Quantum Effects and Applications (SICQEA),
Hunan Normal University, Changsha 410081, China}
	
\begin{abstract}
Inspired by the latest experimental progress, we systematically study the Okubo-Zweig-Iizuka(OZI)-allowed two-body strong decay properties of $1P$-, $1D$-, $2S$- and $2P$-wave $\Lambda_c$ baryons within the $j $-$j$ coupling scheme in the framework of the quark pair creation model. The calculations indicate that: ($\rm{\romannumeral1}$) Taking the observed states $\Lambda_c(2595)^+$ and $\Lambda_c(2625)^+$ as the $1P$-wave $\lambda$-modes states $\Lambda_c|J^P=1/2^-,1\rangle_{\lambda}$ and $\Lambda_c|J^P=3/2^-,1\rangle_{\lambda}$, respectively, we can reproduce the experimental data well in theory. ($\rm{\romannumeral2}$) Combining with the measured mass and the decay properties of $\Lambda_c(2860)^+$, this excited state can be explained as $1D$-wave $\lambda$-mode state $\Lambda_c|J^P=3/2^+,1\rangle_{\lambda\lambda}$. ($\rm{\romannumeral3}$) The newly observed state $\Lambda_c(2910)^+$ may be assigned as one of the $1P$-wave $\rho$-mode states $\Lambda_c|J^P=3/2^-,2\rangle_{\rho}$ or $\Lambda_c|J^P=5/2^-,2\rangle_{\rho}$. Meanwhile, we notice that the partial decay width ratio between $\Sigma_c\pi$ and $\Sigma_c^*\pi$ for the two candidates is significantly different. Hence, experimental progress in this ratio measurement may shed light on the nature of $\Lambda_c(2910)^+$. ($\rm{\romannumeral4}$) According to the properties of $\Lambda_c(2765)^+$, we find that the $2S$-wave $\lambda$-mode state $\Lambda_{c1}|J^P=1/2^+,0\rangle_{\lambda}$ may be a potential candidate. ($\rm{\romannumeral5}$) The $2P$-wave $\lambda$-mode state $\Lambda_{c1}|J^P=3/2^-,1\rangle_{\lambda}$ is mostly likely to be a good assignment of the controversial state $\Lambda_c(2940)^+$. Both the total decay width and partial decay ratio between $pD^0$ and $\Sigma_c\pi$ are in good agreement with the observations. ($\rm{\romannumeral6}$) In addition, for the missing $\Lambda_c$ excitations, we obtain their strong decay properties and hope that's useful for future experimental exploration.
\end{abstract}
	\maketitle

\section{Introduction}

The singly-charmed baryons contain a charm quark and two light quarks, which provide a good opportunity to study the dynamics of quark confinement~\cite{Chen:2016spr,Cheng:2021qpd}. Singly-Charmed baryon spectroscopy has always been a hot topic at the forefront. So far, especially in the last six years, a great progress has been achieved in experiment and many new excited singly-charmed baryons have been discovered~\cite{Chen:2022asf}. To decode the inner structures of these newly observed states, many efforts have been made in both experiment and theory.

The $\Lambda_c$ baryon spectrum is one of important members of the singly-charmed baryons, and there have accumulated some valuable data in experiment. According to the PDG 2024~\cite{ParticleDataGroup:2024cfk}, there are eight $\Lambda_c$ baryons: $\Lambda_c^+$, $\Lambda_c(2595)^+/\Lambda_c(2625)^+$, $\Lambda_c(2765)^+$(or $\Sigma_c(2765)^+$), $\Lambda_c(2860)^+/\Lambda_c(2880)^+$, $\Lambda_c(2910)^+$ and $\Lambda_c(2940)^+$. The $\Lambda_c^+$ ground state is the lowest-lying charmed baryon. It was observed firstly by Fermilab in 1976~\cite{Knapp:1976qw}. $\Lambda_c(2595)^+$ and $\Lambda_c(2625)^+$ are the $\Lambda_c^+$ orbital excitations. They were first reported by the CLEO Collaboration in 1995~\cite{CLEO:1994oxm} and ARGUS Collaboration in 1993~\cite{ARGUS:1993vtm}, respectively, and soon confirmed by subsequent experiments~\cite{CLEO:1994oxm,E687:1993bax,E687:1995srl,ARGUS:1997snv}. The spin-parity of $\Lambda_c(2595)^+$ is almost certainly $1/2^-$, and that of $\Lambda_c(2625)^+$ is expected to be $3/2^-$~\cite{ParticleDataGroup:2024cfk}. The $\Lambda_c(2765)^+$ is a rather broad structure first reported in the $\Lambda_c^+\pi^+\pi^-$ channel by the CLEO Collaboration in 2001~\cite{CLEO:2000mbh} and later also observed in the $\Sigma_c^{++/0}\pi^{\mp}$ decay by the Belle Collaboration in 2007~\cite{Belle:2006xni}. However, nothing at all is known about its quantum numbers, including whether it is a $\Lambda_c^+$ or a $\Sigma_c^+$. In 2017, the Belle Collaboration determined its isospin to be zero, and suggested this particle to be a $\Lambda_c^+$ state~\cite{Belle:2019bab}. The $\Lambda_c(2860)^+$ resonance of spin-parity $3/2^+$ was reported by the LHCb Collaboration in the $D^0p$ amplitude in 2017~\cite{LHCb:2017jym}, and expected to be confirmed by other experiment in the future. Another orbital excitation $\Lambda_c(2880)^+$ of spin-parity $5/2^+$ was first observed by the CLEO Collaboration~\cite{CLEO:2000mbh}. The mass, decay width and quantum number were further determined by later experiments~\cite{Belle:2006xni,LHCb:2017jym,Belle:2024cmc,BaBar:2006itc}. In addition, the BaBar Collaboration also reported a new structure $\Lambda_c(2940)^+$ in the same paper~\cite{BaBar:2006itc}, and soon the Belle Collaboration confirmed the existence of this state and reported first observation of $\Lambda_c(2940)^+\rightarrow \Sigma_c(2455)^{0,++}\pi^{+,-}$ decay~\cite{Belle:2006xni}. Furthermore, the LHCb Collaboration confirmed this structure in the $D^0p$ amplitude~\cite{LHCb:2017jym}, and suggest its spin-parity to be $3/2^-$. Lately, the Belle Collaboration measured the branching fractions of $\Lambda_c(2940)^+$ decaying to $\Lambda_c^+\eta$ and $D^0p$ relative to $\Sigma_c\pi$~\cite{Belle:2024cmc}, which provided a more accurate reference for the theory. The new resonance $\Lambda_c(2910)^+$ was observed by the Belle Collaboration in $\bar{B}^0\rightarrow \Sigma_c(2455)^{0,++}\pi^{\pm}\bar{p}$ decays in 2023~\cite{Belle:2022hnm}, and its spin-parity wasn't be determined now.

Meanwhile, there exist many theoretical calculations to decode the inner structures of those observed $\Lambda_c^+$ baryons via mass spectrum~\cite{Shah:2016mig,Baccouche:2001snw,Baccouche:2001yew,Ebert:2007nw,Weng:2024roa,Chen:2016phw,Yang:2023fsc,Chen:2016iyi,Yao:2018jmc,
Chen:2017aqm,Ebert:2011kk,Yoshida:2015tia,Capstick:1986ter,Chen:2014nyo,Roberts:2007ni,Garcia-Tecocoatzi:2022zrf} and decay properties~\cite{Cho:1994vg,Garcia-Tecocoatzi:2022zrf,Tawfiq:1998nk,Ivanov:1999bk,Tawfiq:1999cf,Chow:1995nw,Pirjol:1997nh,Baccouche:2001snw,
Blechman:2003mq,Gong:2021jkb,Guo:2019ytq,Cheng:2006dk,Nagahiro:2016nsx,Zhong:2007gp,Lu:2018utx,Chen:2007xf}.
Except for the $\Lambda_c^+$ ground state, the $\Lambda_c(2595)^+$ and $\Lambda_c(2625)^+$ can be well interpreted as the $P$-wave $\Lambda_c$ states of $J^P=1/2^-$ and $J^P=3/2^-$~\cite{Shah:2016mig,Baccouche:2001snw,Baccouche:2001yew,Cho:1994vg,Tawfiq:1998nk,Ivanov:1999bk,Tawfiq:1999cf,Chow:1995nw,Pirjol:1997nh}, respectively. Meanwhile, for the $\Lambda_c(2860)^+$ and $\Lambda_c(2880)^+$, most of the references suggest they together form the $D$-wave $\Lambda_c$ doublet of $J^P=3/2^+$ and $J^P=5/2^+$, respectively~\cite{Chen:2016iyi,Chen:2016phw,Yao:2018jmc,Chen:2017aqm,Gong:2021jkb,Guo:2019ytq}. However, there is some literature that don't support $\Lambda_c(2880)^+$ as $J^P=5/2^+$ state~\cite{Chen:2017aqm,Gong:2021jkb}, and believe this state being $F$-wave $\Lambda_c$ states of $J^P=5/2^-$~\cite{Gong:2021jkb}. Hence, the relation between those two states still needs to be carefully examined in future experimental and theoretical studies. As for $\Lambda_c(2940)^+$, its internal structure is controversial. In addition to being interpreted as a traditional hadron state of $J^P=1/2^-$~\cite{Cheng:2021qpd}, $3/2^{\pm}$~\cite{Weng:2024roa,Yang:2023fsc,Gong:2021jkb,Cheng:2006dk}, $5/2^{\mp}$~\cite{Cheng:2006dk,Zhong:2007gp} or $7/2^+$~\cite{Nagahiro:2016nsx}, it has also been interpreted as $D^*N$ molecular state with $J^P=3/2^-$~\cite{Luo:2022cun,Yue:2024paz}. Meanwhile, there are some references~\cite{Xie:2015zga,Ozdem:2023eyz} discussing the properties of $\Lambda_c(2940)$ for helping us to clarify its nature. Fortunately, the latest experimental measurements by Belle Collaboration~\cite{Belle:2024cmc} about the partial decay width ratios of $\Lambda_c(2940)^+$ will provide a stronger basis for decoding its inner structure. Compared to $\Lambda_c(2940)^+$, the properties of $\Lambda_c(2765)^+$(or $\Sigma_c(2765)^+$) and $\Lambda_c(2910)^+$ are more controversial, and it is not even certain whether they are $\Lambda_c$ or $\Sigma_c$ states. At present, theoretical explanations suggest that $\Lambda_c(2765)^+$(or $\Sigma_c(2765)^+$) may be $\Lambda_c(2S)1/2^+$~\cite{Ebert:2007nw,Weng:2024roa}, $\Lambda_c(1P)1/2^-$~\cite{Weng:2024roa} or $\Sigma_c(1P)3/2^-$~\cite{Ebert:2007nw} resonance. For $\Lambda_c(2910)^+$, it can be assigned as $\Lambda_c$ resonance with spin-parity $1/2^+$~\cite{Weng:2024roa} or $1/2^-$~\cite{Yang:2023fsc}. Moreover, the assignment of the two states as $D^{(*)}N$ molecular states also exists~\cite{Yue:2024paz,Montesinos:2024eoy}.

\begin{table*}
	\caption{Predicted masses of $\Lambda_c$ states($1P$-, $1D$-, $2S$- and $2P$-wave) in various quark models and quark pair creation modes.}	
	\label{table1}
	\begin{tabular}{cccccccccccccccc}
		\toprule[1.2pt]
\multirow{1}{*}{Notation}& \multicolumn{8}{c}{Quantum Number}& \multicolumn{5}{c}{Mass} & \multicolumn{1}{c}{Decay channel}  \\\hline
		$\Lambda_c$${\ket{J^P,j}}_{\lambda(\rho)}$&$n_{\lambda}$&$n_{\rho}$&$l_{\lambda}$&$l_{\rho}$&$L$&$s_{\rho}$&$j$&$J^P$
&RQM~\cite{Ebert:2011kk}&RQM~\cite{Capstick:1986ter}&RFT~\cite{Chen:2014nyo}&NCQM~\cite{Chen:2016iyi}&QM~\cite{Roberts:2007ni}& \\  \hline
$\Lambda_c$${\ket{J^P=\frac{1}{2}^-,1}}_\lambda$&0&0&1&0&1&0&1&$\frac{1}{2}^{-}$ &$2598 $&$2630$&$2591$&2614&$2625$ &$\Sigma_c\pi$\\
$\Lambda_c$${\ket{J^P=\frac{3}{2}^-,1}}_\lambda$&0&0&1&0&1&0&1&$\frac{3}{2}^{-}$ &2627      &2640&2629&2639&2636 &\\  \hline
$\Lambda_c$${\ket{J^P=\frac{1}{2}^-,0}}_\rho$&0&0&0&1&1&1&0&$\frac{1}{2}^{-}$ &      &2780&&&2816 &$\Sigma^{(*)}_c\pi$\\
$\Lambda_c$${\ket{J^P=\frac{1}{2}^-,1}}_\rho$&0&0&0&1&1&1&1&$\frac{1}{2}^{-}$ &      &2830&&&2816 &\\
$\Lambda_c$${\ket{J^P=\frac{3}{2}^-,1}}_\rho$&0&0&0&1&1&1&1&$\frac{3}{2}^{-}$ &      &2840&&&2830 &\\
$\Lambda_c$${\ket{J^P=\frac{3}{2}^-,2}}_\rho$&0&0&0&1&1&1&2&$\frac{3}{2}^{-}$ &      &2885&&&2830 &\\
$\Lambda_c$${\ket{J^P=\frac{5}{2}^-,2}}_\rho$&0&0&0&1&1&1&2&$\frac{5}{2}^{-}$ &      &2900&&&2872 &\\ \hline
$\Lambda_c$${\ket{J^P=\frac{3}{2}^+,2}}_{\lambda\lambda}$&0&0&2&0&2&0&2&$\frac{3}{2}^{+}$ &$2874 $&$2910$&$2857$&2843&$2887$ &$\Sigma^{(*)}_c\pi$,$DN$\\
$\Lambda_c$${\ket{J^P=\frac{5}{2}^+,2}}_{\lambda\lambda}$&0&0&2&0&2&0&2&$\frac{5}{2}^{+}$ &2880    &2910&2879&2851&2887 &\\  \hline
$\Lambda_c$${\ket{J^P=\frac{3}{2}^+,2}}_{\rho\rho}$&0&0&0&2&2&0&2&$\frac{3}{2}^{+}$ &&$\sim3035$&&&$3073$ &$\Lambda^{+}_c\omega^0$,$\Sigma^{(*)}_c\pi$,$\Xi^{'}_cK$,$\Sigma_c(1P_{\lambda})\pi$\\
$\Lambda_c$${\ket{J^P=\frac{5}{2}^+,2}}_{\rho\rho}$&0&0&0&2&2&0&2&$\frac{5}{2}^{+}$ &&$\sim3140$&&&3092 &\\  \hline
$\Lambda_{c1}$${\ket{J^P=\frac{1}{2}^+,0}}_\lambda$&1&0&0&0&0&0&0&$\frac{1}{2}^{+}$ &$2769 $&$2775$&$2766$&2772&$2791$ &$\Sigma^{(*)}_c\pi$,$DN$\\ \hline
$\Lambda_{c1}$${\ket{J^P=\frac{1}{2}^+,0}}_\rho$&0&1&0&0&0&0&0&$\frac{1}{2}^{+}$ &       &$2970$&&& &$\Sigma^{(*)}_c\pi$,$D^{(*)}N$,$\Sigma_c(1P_{\lambda})\pi$\\ 	\hline	
$\Lambda_{c1}$${\ket{J^P=\frac{1}{2}^-,1}}_\lambda$&1&0&1&0&1&0&1&$\frac{1}{2}^{-}$ &$2983 $&$3030$&$2989$&2980& &$\Sigma^{(*)}_c\pi$,$D^{(*)}N$,$\Sigma_c(1P_{\lambda})\pi$\\
$\Lambda_{c1}$${\ket{J^P=\frac{3}{2}^-,1}}_\lambda$&1&0&1&0&1&0&1&$\frac{3}{2}^{-}$ &$3005 $&$3035$&$3000$&3004& &\\  \hline
$\Lambda_{c1}$${\ket{J^P=\frac{1}{2}^-,0}}_\rho$&0&1&0&1&1&1&0&$\frac{1}{2}^{-}$ &       &$3200$&&& &$\Lambda^{+}_c\omega^0/\eta$,$\Sigma^{(*)}_c\pi$,$\Xi_cK$,$\Xi_c'^{(*)}K$,$\Sigma_c(1P_{\lambda/\rho})\pi$,
$\Lambda^{+}_c(1P_{\lambda})\eta$\\
$\Lambda_{c1}$${\ket{J^P=\frac{1}{2}^-,1}}_\rho$&0&1&0&1&1&1&1&$\frac{1}{2}^{-}$ &       &$3240$&&& &\\
$\Lambda_{c1}$${\ket{J^P=\frac{3}{2}^-,1}}_\rho$&0&1&0&1&1&1&1&$\frac{3}{2}^{-}$ &       &$3240$&&& &\\
$\Lambda_{c1}$${\ket{J^P=\frac{3}{2}^-,2}}_\rho$&0&1&0&1&1&1&2&$\frac{3}{2}^{-}$ &       &$3255$&&& &\\
$\Lambda_{c1}$${\ket{J^P=\frac{5}{2}^-,2}}_\rho$&0&1&0&1&1&1&2&$\frac{5}{2}^{-}$ &       &$3130$&&& &\\
\bottomrule[1.2pt]
	\end{tabular}
\end{table*}

To decode the inner structures of those undetermined $\Lambda_c$ resonances, more theoretical and experimental efforts are essential. Meanwhile, the studies on strong decay properties of $\rho$-mode excitations are scarce. Hence,
in the present work, we carry out a systematic analysis of $1P$-, $1D$-, $2S$- and $2P$-wave $\Lambda_c$ states for both $\rho$- and
$\lambda$-mode excitations within the quark pair creation model. On the one hand, we attempt to explain properties of the controversial states, and on the other hand, we want to predict the decays for unobserved $\Lambda_c$ states. The predicted masses and possible decay channels within the quark pair creation model are collected in Table~\ref{table1}.

This paper is structured as follows. In Sec. II, we briefly
introduce the quark pair creation model.
Then we present our theoretical results and discussions in Sec. III. A summary is given in Sec. IV.

\section{Theoretical framework}

\begin{figure}[h]
	\begin{center}
		\subfigure[]{\begin{minipage}{0.32\linewidth}
				\centering
				\includegraphics[width=1in,height=1in]{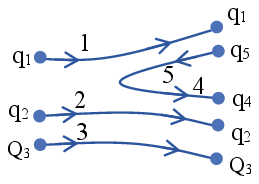}
		\end{minipage}}
		\subfigure[]{\begin{minipage}{0.32\linewidth}
				\centering
				\includegraphics[width=1in,height=1in]{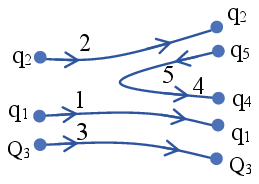}
		\end{minipage}}
		\subfigure[]{\begin{minipage}{0.32\linewidth}
				\centering
				\includegraphics[width=1in,height=1in]{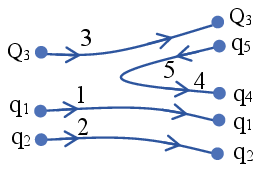}
		\end{minipage}}
		\centering
		\caption{Possible decay ways for the $\Lambda_c$ system within the quark pair creation model.}\label{FIG1}
	\end{center}
\end{figure}

The quark pair creation model~\cite{Micu:1968mk,Carlitz:1970xb,LeYaouanc:1972vsx,LeYaouanc:1977fsz,LeYaouanc:1977gm} as a phenomenological method has been employed successfully in the description of the OZI-allowed two-body strong decays.
The main idea of this model is that the quark-antiquark pair with $0^{++}$ is created from the vacuum and then regroups with the quarks from the initial hadron to produce two outing hadrons. Hence, for the $\Lambda_c$ system, there are three decay processes, as shown in Fig~\ref{FIG1}.

In the framework of the quark pair creation model, the transition operator for a two-body decay($A\rightarrow B+C$) in the nonrelativistic limit reads
\begin{equation}
	\begin{aligned}
	T=&-3\gamma\sum\limits_{m}\langle1m;1-m|00\rangle\int d^3\textbf{p}_4d^3\textbf{p}_5\delta^3(\textbf{p}_4+\textbf{p}_5)\\ &\times\omega^{45}_0\chi^{45}_{1,-m}\phi^{45}_0\mathcal{Y}^m_1(\dfrac{\textbf{p}_4-\textbf{p}_5}{2})a^\mathcal{y}_{4i}(\textbf{p}_4)b^\mathcal{y}_{5j}(\textbf{p}_5).\\
\end{aligned}
\end{equation}
The pair creation strength $\gamma$ is a dimensionless parameter and fixed by fitting experimental data. $\textbf{p}_i(i=4,5)$ denotes the three-vector momentum of the $i$th quark of the created quark pair. $\omega^{45}_0$=$\delta_{ij}$ and $\chi^{45}_{1,-m}$ stand for the color singlet and spin triplet of the quark pair, respectively. $\phi^{45}_0$=(u$\rm{\overline{u}}$+d$\rm{\overline{d}}$+s$\rm{\overline{s}}$)/$\sqrt{3}$ represents the flavor function. The solid harmonic polynomial $\mathcal{Y}^m_1$=$|\textbf{p}|$$\rm{\boldsymbol{Y}}^m_1$($\theta_p\phi_p$) corresponds to the momentum-space distribution. The creation operator $a^\mathcal{y}_{4i}d^\mathcal{y}_{5j}$ is the quark pair-creation in the vacuum.

According to the definition of the mock state~\cite{Hayne:1981zy}, the wave
functions of the baryon(denoted as $|A\rangle$) and meson(denoted as $|C\rangle$) is given by, respectively,
\begin{eqnarray}
|A(N_A~^{2S_A+1}L_A{J_AM_{J_A}})(\mathbf{p}_A)\rangle=~~~~~~~~~~~~~~~~~~~~~~~~~~~~~~~~~~~~~~\nonumber \\
\sqrt{2E_A}\varphi^{123}_A\omega^{123}_A\sum_{M_{L_A},M_{S_A}}\langle L_AM_{L_A};S_AM_{S_A}|J_AM_{J_A}\rangle\nonumber \\
\times\int d^3\mathbf{p}_1d^3\mathbf{p}_2d^3\mathbf{p}_3\delta^3(\mathbf{p}_1+\mathbf{p}_2+\mathbf{p}_3-\mathbf{p}_A)\nonumber \\
\times\Psi_{N_AL_AM_{L_A}(\mathbf{p}_1,\mathbf{p}_2,\mathbf{p}_3)}\chi^{123}_{S_AM_{S_A}}
|q_1(\mathbf{p}_1)q_2(\mathbf{p}_2)q_3(\mathbf{p}_3)\rangle,
\end{eqnarray}
\begin{eqnarray}
|C(N_C~^{2S_C+1}L_C{J_CM_{J_C}})(\mathbf{p}_C)\rangle=~~~~~~~~~~~~~~~~~~~~~~~~~~~~~~~~~~~~~~\nonumber \\
\sqrt{2E_C}\varphi^{ab}_C\omega^{ab}_C\sum_{M_{L_C},M_{S_C}}\langle L_CM_{L_C};S_CM_{S_C}|J_CM_{J_C}\rangle\nonumber \\
\times\int d^3\mathbf{p}_ad^3\mathbf{p}_b\delta^3(\mathbf{p}_a+\mathbf{p}_b-\mathbf{p}_C)\nonumber \\
\times\Psi_{N_CL_CM_{L_C}(\mathbf{p}_a,\mathbf{p}_b)}\chi^{ab}_{S_CM_{S_C}}|q_a(\mathbf{p}_a)q_b(\mathbf{p}_b)\rangle.
\end{eqnarray}
The $\textbf{p}_i$ ($i = 1, 2, 3$ and $a, b$) stands for the momentum of quarks
in baryon $|A\rangle$ and meson $|C\rangle$. $\textbf{p}_A$($\textbf{p}_C$) denotes the momentum of the hadron $|A\rangle$($|C\rangle$).
The spatial wave functions of hadrons are described with simple harmonic oscillator wave functions. For a baryon without the radial excitation reads
\begin{equation}
\begin{split}
&\psi(l_{\rho},m_{\rho},l_{\lambda},m_{\lambda})\\
&=3^{\frac{3}{4}}(-i)^{l_{\rho}}\left[\dfrac{2^{l_{\rho}+2}}{\sqrt{\pi}(2l_{\rho}+1)!!}\right]^\frac{1}{2}
\left(\dfrac{1}{\alpha_{\rho}}\right)^{l_{\rho}+\frac{3}{2}}{\rm {exp}}(-\frac{\textbf{p}_{\rho}^2}{2\alpha_{\rho}^2})\mathcal{Y}^{m_{\rho}}_{l_{\rho}}(\textbf{p}_{\rho})\\
&\times(-i)^{l_{\lambda}}\left[\dfrac{2^{l_{\lambda}+2}}{\sqrt{\pi}(2l_{\lambda}+1)!!}\right]^\frac{1}{2}
\left(\dfrac{1}{\alpha_{\lambda}}\right)^{l_{\lambda}+\frac{3}{2}}{\rm {exp}}(-\frac{\textbf{p}_{\lambda}^2}{2\alpha_{\lambda}^2})\mathcal{Y}^{m_{\lambda}}_{l_{\lambda}}(\textbf{p}_{\lambda}),
\end{split}
\end{equation}
and for a baryon with one radial excitation($n_{\lambda/\rho}=1$) reads
\begin{equation}
\begin{split}
&\psi^{n_{\lambda/\rho}=1}(l_{\rho},m_{\rho},l_{\lambda},m_{\lambda})\\
&=3^{\frac{3}{4}}(-i)^{l_{\lambda/\rho}+2}\left[\dfrac{2^{l_{\lambda/\rho}+3}}{\sqrt{\pi}(2l_{\lambda/\rho}+3)!!}\right]^\frac{1}{2}
\left(\dfrac{1}{\alpha_{\lambda/\rho}}\right)^{l_{\lambda/\rho}+\frac{3}{2}}{\rm {exp}}(-\frac{\textbf{p}_{\lambda/\rho}^2}{2\alpha_{\lambda/\rho}^2})\\
&\times\mathcal{Y}^{m_{\lambda/\rho}}_{l_{\lambda/\rho}}(\textbf{p}_{\lambda/\rho})
\left(\frac{2l_{\lambda/\rho}+3}{2}-\frac{\textbf{p}_{\lambda/\rho}^2}{\alpha_{\lambda/\rho}^2}\right)\\
&\times(-i)^{l_{\rho/\lambda}}\left[\dfrac{2^{l_{\rho/\lambda}+2}}{\sqrt{\pi}(2l_{\rho/\lambda}+1)!!}\right]^\frac{1}{2}
\left(\dfrac{1}{\alpha_{\rho/\lambda}}\right)^{l_{\rho/\lambda}+\frac{3}{2}}{\rm {exp}}(-\frac{\textbf{p}_{\rho/\lambda}^2}{2\alpha_{\rho/\lambda}^2})\mathcal{Y}^{m_{\rho/\lambda}}_{l_{\rho/\lambda}}(\textbf{p}_{\rho/\lambda}).
\end{split}
\end{equation}
The $\textbf{p}_{\rho}$ represents the relative momentum within the light diquark, and $\textbf{p}_{\lambda}$ denotes the relative momentum between the light diquark and the heavy quark.
The spatial wave function for a ground meson $|C\rangle$ is
\begin{equation}
	\psi_{0,0}=\left(\dfrac{R^2}{\pi}\right)^\frac{3}{4}{\rm {exp}}\left(-\dfrac{R^2\textbf{p}^2_{ab}}{2}\right),
\end{equation}
where the $\textbf{p}_{ab}$ is the relative momentum between the
quark and antiquark in the meson.

Then, in the center of mass frame we can obtain the partial decay amplitude,
\begin{equation}
	\begin{aligned}
		&\mathcal{M}^{M_{J_A}M_{J_B}M_{J_C}}(A\rightarrow B+C)\\
		=&\gamma\sqrt{8E_AE_BE_C}\prod_{A,B,C}\langle\chi^{124}_{S_BM_{S_B}}\chi^{35}_{S_CM_{S_C}}|\chi^{123}_{S_AM_{S_A}}\chi^{45}_{1-m}\rangle\\
		&\langle\varphi^{124}_B\varphi^{35}_C|\varphi^{123}_A\varphi^{45}_0\rangle I^{M_{L_A},m}_{M_{L_B},M_{L_C}}(\textbf{p}).
	\end{aligned}
\end{equation}
$I^{M_{L_A},m}_{M_{L_B},M_{L_C}}(\textbf{p})$ denotes the spatial integration, and Clebsch-Gorden coefficient $\prod_{A,B,C}$ reads
\begin{equation}
	\begin{aligned}
			\sum&\langle L_BM_{L_B};S_BM_{S_B}|J_B,M_{J_B}\rangle \langle L_CM_{L_C};S_CM_{S_C}|J_C,M_{J_C}\rangle\\
			 &\times \langle L_AM_{L_A};S_AM_{S_A}|J_A,M_{J_A}\rangle \langle1m;1-m|00\rangle.\\
	\end{aligned}
\end{equation}

Considering the vertex given by the quark pair creation model too hard at high momenta, we modify the vertex by adopting a form factor $e^{-\frac{\mathbf{p}^2}{2\Lambda^2}}$ as in the literatures~\cite{Ortega:2016mms,Morel:2002vk,Ortega:2016pgg}, which gives the quark-pair-creation vertex a finite-size rather than point-like behavior. It reads
\begin{equation}
	\begin{aligned}
		\mathcal{M}^{M_{J_A}M_{J_B}M_{J_C}}(A\rightarrow B+C)\rightarrow \mathcal{M}^{M_{J_A}M_{J_B}M_{J_C}}(A\rightarrow B+C)e^{-\frac{\mathbf{p}^2}{2\Lambda^2}}.
	\end{aligned}
\end{equation}
In the equation, we fix the cut-off parameter $\Lambda=780$ MeV, which is the same as to the value used in Ref.~\cite{Ni:2023lvx}. $\textbf{p}$ stands for the momentum of the daughter baryon $B$ in the center of mass frame of the parent baryon $A$, and reads
\begin{equation}
	|\textbf{p}|=\dfrac{\sqrt{[M_A^2-(M_B-M_C)^2][M_A^2-(M_B+M_C)^2]}}{2M_A}.
\end{equation}

Finally, the decay width $\Gamma[A\rightarrow BC]$ can be calculated by the following formula,
\begin{equation}
\Gamma(A\rightarrow BC)=\pi^2 \dfrac{|\textbf{p}|}{M^2_A}\dfrac{1}{2J_A+1}\sum_{M_{J_A},M_{J_B},M_{J_C}}|\mathcal{M}^{M_{J_A},M_{J_B},M_{J_C}}|^2.
\end{equation}

\begin{table}[htpb]
\caption{\label{finalstates} Masses (MeV) of the final baryons and mesons~\cite{ParticleDataGroup:2024cfk,Yu:2022ymb,Yoshida:2015tia}. }
\begin{tabular}{ccccccccccc}\hline\hline
State~~~~&Mass~~~~&State~~~~&Mass~~~~&State~~~~&Mass \\
$p$~~~~&938.27~~~~~&$\pi^0$~~~~&134.98~~~~&$\eta'$ &957.78 \\
$n$~~~~&939.57~~~~~&$\pi^{\pm}$~~~~&139.57~~~~&$\rho$ &775.26 \\
$\Lambda_c^+$~~~~&2286.46~~~~&$\eta$~~~~&547.862~~~~&$\Lambda_c|J^{P}=\frac{1}{2}^-,1\rangle_{\lambda}$ ~~~~&2592.25 \\
$\Sigma_c^0$~~~~&2453.75~~~~&$\omega$~~~~&782.66~~~~&$\Lambda_c|J^{P}=\frac{3}{2}^-,1\rangle_{\lambda}$ ~~~~&2628.00 \\
$\Sigma_c^+$~~~~&2452.65~~~~&$K^0$~~~~&497.611~~~~&$\Sigma_c|J^{P}=\frac{1}{2}^-,0\rangle_{\lambda}$ ~~~~&2823 \\
$\Sigma_c^{++}$~~~~&2453.97~~~~&$K^+$~~~~&493.677~~~~&$\Sigma_c|J^{P}=\frac{1}{2}^-,1\rangle_{\lambda}$ ~~~~&2809 \\
$\Xi_c^+$~~~~&2467.71~~~~&$K^{0*}$~~~~&895.55~~~~&$\Sigma_c|J^{P}=\frac{3}{2}^-,1\rangle_{\lambda}$ ~~~~&2829 \\
$\Xi_c^0$~~~~&2470.44~~~~&$K^{+*}$~~~~&891.67~~~~&$\Sigma_c|J^{P}=\frac{3}{2}^-,2\rangle_{\lambda}$ ~~~~&2802 \\
$\Xi_c'^{+}$~~~~&2578.2~~~~&$D^0$~~~~&1864.84~~~~&$\Sigma_c|J^{P}=\frac{5}{2}^-,2\rangle_{\lambda}$ ~~~~&2835 \\
$\Xi_c'^{0}$~~~~&2578.7~~~~&$D^+$~~~~&1869.66~~~~&$\Sigma_c|J^{P}=\frac{1}{2}^-,1\rangle_{\rho}$ ~~~~&2909\\
$\Xi_c'^{+*}$~~~~&2645.1~~~~&$D^{0*}$~~~~&2006.85~~~~&$\Sigma_c|J^{P}=\frac{3}{2}^-,1\rangle_{\rho}$ ~~~~&2910 \\
$\Xi_c'^{0*}$~~~~&2646.16~~~~&$D^{+*}$~~~~&2010.26~~~~& & \\
\hline\hline
\end{tabular}
\end{table}

In this work, we adopt $m_u$=$m_d$=330 MeV, $m_s$=450 MeV, and $m_c$=1700 MeV for the constituent quark mass. The masses of the final baryons and mesons involved in our calculations, collected in Table~\ref{finalstates}. The harmonic oscillator strength $R=2.5$ GeV$^{-1}$ for light flavor mesons $\pi/K^{(*)}/\omega/\eta$, $R=1.67$ GeV$^{-1}$ for $D$ meson and $R=1.94$ GeV$^{-1}$ for $D^*$ meson~\cite{Godfrey:2015dva}. The parameter of the $\rho$-mode excitation between the two light quarks is taken as $\alpha_{\rho}$=0.4 GeV. The other parameter $\alpha_\lambda$ is obtained by the relation~\cite{Zhong:2007gp}
\begin{equation}
	\alpha_\lambda=\left(\dfrac{3m_Q}{2m_q+m_Q}\right)^\frac{1}{4}\alpha_\rho.
\end{equation}
The value of vacuum pair-production strength $\gamma$ is determined by fitting the well measured decay $\Sigma_c^{++}(2520)\rightarrow \Lambda_c^+\pi^+$, and is obtained as $\gamma=$11.51.

\section{Calculations and Results }

The two-body strong decays of
 $1P$-, $1D$-, $2S$- and $2P$-wave excited $\Lambda_c$ baryons within the $j$-$j$ coupling scheme are systematically investigated by
the quark pair creation model. Both $\lambda$-mode and $\rho$-mode excitations are considered in this calculation.
We attempt to decode the inner structures of the observed controversial $\Lambda_c$ excitations, while give the strong decay predictions for the missing $\Lambda_c$ states, which may be helpful for the observation in forthcoming experiments.

\subsection{$1P$-wave $\lambda$-mode excitations}

For the $1P$-wave $\lambda$-mode $\Lambda_c$ baryons, there are two states according to the quark model classification, which are $\Lambda_c|J^P=1/2^-,1\rangle_{\lambda}$ and $\Lambda_c|J^P=3/2^-,1\rangle_{\lambda}$. They correspond to the well determined states $\Lambda_c(2595)^+$ and $\Lambda_c(2625)^+$, respectively. Hence, we fix the masses of the two states at the physical masses, and collect their decay properties in Table~\ref{Plambda}.

\begin{table}[h]
	\caption{\label{Plambda}The strong decay properties of the $\lambda$-mode $1P$-wave $\Lambda_c$ states, which are taken as $\Lambda_c(2595)^+$ and $\Lambda_c(2625)^+$, respectively. $\Gamma_{\text{Total}}$ represents the total decay width and Expt. denotes the experimental value. The unit is MeV.}
	\centering
	\begin{tabular}{c|c|ccccccccccccccc}	
		\hline\hline
\multirow{2}{*}{Decay width}	&$\Lambda_{c}\ket{J^{P}=\frac{1}{2}^{-},1}_{\lambda}$ &$\Lambda_{c}\ket{J^{P}=\frac{3}{2}^{-},1}_{\lambda}$\\ \cline{2-3}
                                &$\Lambda_c(2595)^+$&$\Lambda_c(2625)^+$\\ \hline
 $\Gamma[\Sigma_c^0\pi^+]$     &--&0.02  \\
$\Gamma[\Sigma_c^+\pi^0]$     &7.07&0.01 \\
$\Gamma[\Sigma_c^{++}\pi^-]$     &--&0.01   \\ \hline
$\Gamma_{\mathrm{Total}}$     &7.07& 0.04 \\  \hline
Expt.     &$2.59\pm0.30\pm0.47$&$<0.52$  \\  \hline\hline
\end{tabular}	
\end{table}

Considering the uncertainties for the experimental data and theoretical calculations, the theoretical value is roughly consistent with the observations, which proves the applicability of the quark pair creation model. In addition, it should be mentioned that the mass of $\Lambda_c(2595)^+$ is very close to the threshold of $\Sigma_c\pi$, and the partial decay widths are highly sensitive to the precision of mass.

\subsection{$1P$-wave $\rho$-mode excitations}
For the $1P$-wave $\rho$-mode $\Lambda_c$ baryons, there are five states $\Lambda_{c}\ket{J^{P}=1/2^{-},0}_{\rho}$, $\Lambda_{c}\ket{J^{P}=1/2^{-},1}_{\rho}$, $\Lambda_{c}\ket{J^{P}=3/2^{-},1}_{\rho}$, $\Lambda_{c}\ket{J^{P}=3/2^{-},2}_{\rho}$ and $\Lambda_{c}\ket{J^{P}=5/2^{-},2}_{\rho}$. According to the theoretical predictions by various methods, the mass of the $1P$-wave $\rho$-mode $\Lambda_c$ baryons is about $M\sim$2.85 GeV. Meanwhile, we notice that their masses are above the threshold of $ND$, while their strong decays are forbidden due to the orthogonality of spatial wave functions. This is true for all of the $\rho$-mode excitations. Hence, we mainly focus on their strong decays into $\Sigma_c\pi$ and $\Sigma_c^*\pi$. Fixing the masses of $1P$-wave $\rho$-mode $\Lambda_c$ baryons at the predictions in Ref.~\cite{Capstick:1986ter}, we study their strong decay properties, and list in Table~\ref{Prho}.

\begin{table*}[]
	\caption{\label{Prho}The strong decay properties of the $\rho$-mode $1P$- and $2P$-wave $\Lambda_c$ states within the quark pair creation model, which masses are taken from the predictions in Ref.~\cite{Capstick:1986ter}. $\Gamma_{\text{Total}}$ represents the total decay width and Expt. denotes the experimental value. The unit is MeV.}
	\centering
	\begin{tabular}{c|c|c|c|cc|cc}	
		\hline\hline
\multirow{2}{*}{Decay width}	&\multicolumn{1}{c}{$\Lambda_{c}\ket{J^{P}=\frac{1}{2}^{-},0}_{\rho}$} &\multicolumn{1}{c}{$\Lambda_{c}\ket{J^{P}=\frac{1}{2}^{-},1}_{\rho}$} &\multicolumn{1}{c}{$\Lambda_{c}\ket{J^{P}=\frac{3}{2}^{-},1}_{\rho}$}  &\multicolumn{2}{c}{$\Lambda_{c}\ket{J^{P}=\frac{3}{2}^{-},2}_{\rho}$}  &\multicolumn{2}{c}{$\Lambda_{c}\ket{J^{P}=\frac{5}{2}^{-},2}_{\rho}$}\\ \cline{2-8}
                                &M=2780&M=2830&M=2840&M=2885&$\Lambda_c(2910)^+$&M=2900&$\Lambda_c(2910)^+$    \\ \hline
 $\Gamma[\Sigma_c\pi]$     &0.00&782.44&10.33&31.53&41.96&16.34&16.65  \\
$\Gamma[\Sigma_c^*\pi]$     &0.00&6.42&768.63&14.37&21.12&27.50&32.85 \\
$\Gamma_{\mathrm{Total}}$     &0.00&788.86&778.96&45.90&63.08&43.84&49.50  \\  \hline
Expt. &\multicolumn{1}{c}{-}&\multicolumn{1}{c}{-}&\multicolumn{1}{c}{-}&\multicolumn{4}{c}{$51.8\pm20.0\pm18.8$}\\
\hline\hline
\multirow{2}{*}{Decay width}	&\multicolumn{1}{c}{$\Lambda_{c1}\ket{J^{P}=\frac{1}{2}^{-},0}_{\rho}$} &\multicolumn{1}{c}{$\Lambda_{c1}\ket{J^{P}=\frac{1}{2}^{-},1}_{\rho}$} &\multicolumn{1}{c}{$\Lambda_{c1}\ket{J^{P}=\frac{3}{2}^{-},1}_{\rho}$}  &\multicolumn{2}{c}{$\Lambda_{c1}\ket{J^{P}=\frac{3}{2}^{-},2}_{\rho}$}  &\multicolumn{2}{c}{$\Lambda_{c1}\ket{J^{P}=\frac{5}{2}^{-},2}_{\rho}$}\\ \cline{2-8}
                                &M=3200&M=3240&M=3240&\multicolumn{2}{c|}{M=3255}&\multicolumn{2}{c}{M=3130}    \\ \hline
$\Gamma[\Sigma_c\pi]$     &0.00&19.02&7.10&\multicolumn{2}{c|}{12.15}&\multicolumn{2}{c}{6.78}  \\
$\Gamma[\Sigma_c^*\pi]$     &0.00&16.47&21.08&\multicolumn{2}{c|}{14.47}&\multicolumn{2}{c}{22.55}  \\
$\Gamma[\Lambda_c\omega]$     &0.00&7.30&7.30&\multicolumn{2}{c|}{12.58}&\multicolumn{2}{c}{2.10}  \\
$\Gamma[\Lambda_c\eta]$     &3.61&0.00&0.00&\multicolumn{2}{c|}{3.38}&\multicolumn{2}{c}{4.15}  \\
$\Gamma[\Lambda_c\eta']$     &-&-&-&\multicolumn{2}{c|}{3.38}&\multicolumn{2}{c}{-}  \\
$\Gamma[\Sigma_c\rho]$     &-&0.23&0.11&\multicolumn{2}{c|}{176.79}&\multicolumn{2}{c}{-}  \\
$\Gamma[\Xi_cK]$     &0.13&0.00&0.00&\multicolumn{2}{c|}{2.65}&\multicolumn{2}{c}{1.36}  \\
$\Gamma[\Xi_c'K]$     &0.00&1.86&0.58&\multicolumn{2}{c|}{1.21}&\multicolumn{2}{c}{0.05}  \\
$\Gamma[\Xi_c^*K]$     &0.00&0.44&6.05&\multicolumn{2}{c|}{0.53}&\multicolumn{2}{c}{-}  \\
$\Gamma[\Lambda_c|J^P=1/2^-,1\rangle_{\lambda}\eta]$     &0.00&0.04&0.01&\multicolumn{2}{c|}{1.95}&\multicolumn{2}{c}{-}  \\
$\Gamma[\Lambda_c|J^P=3/2^-,1\rangle_{\lambda}\eta]$     &0.00&0.01&0.03&\multicolumn{2}{c|}{0.25}&\multicolumn{2}{c}{-}  \\
$\Gamma[\Sigma_c|J^P=1/2^-,0\rangle_{\lambda}\pi]$     &0.00&2.58&2.58&\multicolumn{2}{c|}{0.00}&\multicolumn{2}{c}{0.00}  \\
$\Gamma[\Sigma_c|J^P=1/2^-,1\rangle_{\lambda}\pi]$     &0.15&6.64&1.66&\multicolumn{2}{c|}{13.51}&\multicolumn{2}{c}{0.02}  \\
$\Gamma[\Sigma_c|J^P=3/2^-,1\rangle_{\lambda}\pi]$     &0.28&2.77&6.92&\multicolumn{2}{c|}{1.41}&\multicolumn{2}{c}{6.84}  \\
$\Gamma[\Sigma_c|J^P=3/2^-,2\rangle_{\lambda}\pi]$     &0.00&0.24&0.41&\multicolumn{2}{c|}{3.76}&\multicolumn{2}{c}{0.14}  \\
$\Gamma[\Sigma_c|J^P=5/2^-,2\rangle_{\lambda}\pi]$     &0.00&0.40&0.05&\multicolumn{2}{c|}{3.29}&\multicolumn{2}{c}{0.90}  \\
$\Gamma[\Sigma_c|J^P=1/2^-,1\rangle_{\rho}\pi]$     &57.44&10.96&2.74&\multicolumn{2}{c|}{33.77}&\multicolumn{2}{c}{0.01}  \\
$\Gamma[\Sigma_c|J^P=3/2^-,1\rangle_{\rho}\pi]$     &113.60&5.44&14.27&\multicolumn{2}{c|}{7.94}&\multicolumn{2}{c}{6.94}  \\
$\Gamma_{\mathrm{Total}}$     &175.21&74.40&70.89&\multicolumn{2}{c|}{293.02}&\multicolumn{2}{c}{51.84}  \\ \hline\hline
\end{tabular}	
\end{table*}

\begin{figure}[h]
	\centering \epsfxsize=6 cm \epsfbox{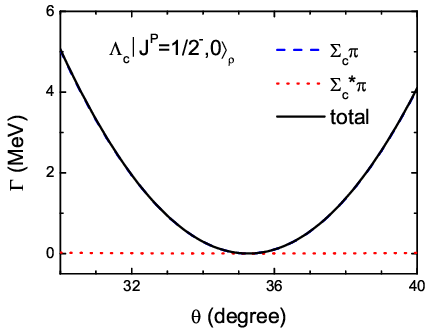}
	\caption{ Partial and total strong decay widths of $\Lambda_c|J^P=1/2^-,0\rangle_{\rho}$ as a function of the mixing angle.}
    \label{FIG2}
\end{figure}

Within the $j-j$ coupling scheme, the total decay width of $\Lambda_c|J^P=1/2^-,0\rangle_{\rho}$ is most likely to be near zero. We know that the states in the $j-j$ coupling scheme can be expressed with the linear combination of the configurations in the $L-S$ coupling scheme, which reads
 \begin{equation}
	\begin{aligned}
		\left|\left\{[(l_{\rho}l_{\lambda})_L s_{\rho}]_j s_Q\right\}_{J^P}
		\right \rangle&=(-1)^{L+s_{\rho}+\frac{1}{2}+J}\sqrt{2J+1}\sum_{S}\sqrt{2S+1}\\
		&\begin{pmatrix}
			L&s_{\rho}&j\\s_Q&J&S
		\end{pmatrix}	\left|\left\{[(l_{\rho}l_{\lambda})_L (s_{\rho}s_Q)_S]_J \right\}
		\right\rangle.
	\end{aligned}
\end{equation}
In the expression, $l_{\rho}$ and $l_{\lambda}$ are the $\rho$- and $\lambda$-modes quantum numbers of the orbital angular, respectively. The total orbital angular momentum $L$ = $|l_{\rho}-l_{\lambda}|,\cdots , l_{\rho}+l_{\lambda} $. $s_{\rho}$ is the quantum numbers of the
total spin of the two light quarks and $s_Q$ is the spin of the
heavy quark. The total spin angular momentum $S$ = $|s_{\rho}-s_Q|,\cdots , s_{\rho}+s_Q $. $J$ is the total angular momentum.
  That means the states in the $j-j$ coupling scheme will contain a mixing angle $\theta$. Considering the heavy quark symmetry being not strictly true and slightly breaking in the $\Lambda_c$ system, the mixing angle $\theta$ will fluctuate around the center value($\theta\simeq35^{\circ}$). To investigate this effect, we plot the strong decay widths of $\Lambda_c|J^P=1/2^-,0\rangle_{\rho}$ as a function of the mixing angle in Fig~\ref{FIG2}. we can obtained that $\Lambda_c|J^P=1/2^-,0\rangle_{\rho}$ is still very narrow state, and the $\Sigma_c\pi$ decay channel almost saturates its total decay widths.

The two states $\Lambda_c|J^P=1/2^-,1\rangle_{\rho}$ and $\Lambda_c|J^P=3/2^-,1\rangle_{\rho}$ are probably broad states with a total decay width around $\Gamma_{\mathrm{Total}}\sim780$ MeV. The $\Lambda_c|J^P=1/2^-,1\rangle_{\lambda}$ dominantly decays into the $\Sigma_c\pi$ channel. While the $\Lambda_c|J^P=3/2^-,1\rangle_{\rho}$ mainly decays into the $\Sigma_c^*\pi$ channel. In this case, the $\Lambda_c|J^P=1/2^-,1\rangle_{\rho}$ and $\Lambda_c|J^P=3/2^-,1\rangle_{\rho}$ might be too broad to observed in experiments.

 The states $\Lambda_c|J^P=3/2^-,2\rangle_{\rho}$ and $\Lambda_c|J^P=5/2^-,2\rangle_{\rho}$ may be moderate states with a total width of several tens MeV, and their strong decays are governed by the $\Sigma_c\pi$ and $\Sigma_c^*\pi$ channels. We notice that the new state $\Lambda_c(2910)^+$ is observed in the $\Sigma_c\pi$ channel by the Belle Collaboration~\cite{Belle:2022hnm}. Its mass and width are measured to be $M=2913.8\pm5.6\pm3.8$ MeV and $\Gamma=51.8\pm20.0\pm18.8$ MeV, respectively. Combining the predicted masses and decay properties, $\Lambda_c|J^P=3/2^-,2\rangle_{\rho}$ and $\Lambda_c|J^P=5/2^-,2\rangle_{\rho}$ may be candidates of $\Lambda_c(2910)^+$. Hence, we fix the masses of the two states as $M=2914$ MeV, and collected their decays in Table~\ref{Prho} as well. It is found that the total decay width of $\Lambda_c|J^P=3/2^-,2\rangle_{\rho}$ is about
 \begin{equation}
 \Gamma_{\mathrm{Total}}\simeq63.08~\mathrm{MeV},
 \end{equation}
 which is consistent with the observation. Meanwhile the main decay channel is $\Sigma_c\pi$ and the predicted partial decay width ratio is
\begin{equation}
\frac{\Gamma[\Lambda_c|J^P=\frac{3}{2}^-,2\rangle_{\rho}]\rightarrow\Sigma_c\pi}{\Gamma[\Lambda_c|J^P=\frac{3}{2}^-,2\rangle_{\rho}\rightarrow\Sigma_c^*\pi]}\simeq1.99.
\end{equation}
This calculation is consistent with the fact that the $\Lambda_c(2910)^+$ was firstly observed in $\Sigma_c\pi$ invariant mass distribution by Belle Collaboration~\cite{Belle:2022hnm}.

As to $\Lambda_c|J^P=5/2^-,2\rangle_{\rho}$, its total decay width is around
 \begin{equation}
 \Gamma_{\mathrm{Total}}\simeq49.50~\mathrm{MeV},
 \end{equation}
which is agreement with the observation as well. The dominant decay channel is $\Sigma_c^*\pi$, and the corresponding branching fraction is
\begin{equation}
\frac{\Gamma[\Lambda_c|J^P=\frac{5}{2}^-,2\rangle_{\rho}]\rightarrow\Sigma_c^*\pi}{\Gamma_{\mathrm{Total}}}\simeq66\%.
\end{equation}
Meanwhile, the predicted branching ratio of the $\Sigma_c\pi$ channel is
\begin{equation}
\frac{\Gamma[\Lambda_c|J^P=\frac{5}{2}^-,2\rangle_{\rho}]\rightarrow\Sigma_c\pi}{\Gamma_{\mathrm{Total}}}\simeq34\%,
\end{equation}
which is enough large to be observed in experiment as well.
To further decode the nature of $\Lambda_c(2910)^+$ and determine whether it's $\Lambda_c|J^P=3/2^-,2\rangle_{\rho}$ or $\Lambda_c|J^P=5/2^-,2\rangle_{\rho}$, the partial decay width ratio between $\Sigma_c\pi$ and $\Sigma_c^*\pi$ may be a good criterion.

Of course, considering that the current experimental data are limited and bare large errors, there are other possible explanations, such as $D^*N$ molecular state~\cite{Yue:2024paz,Montesinos:2024eoy} or $\Lambda_c$ resonances with different spin-parity~\cite{Weng:2024roa,Yang:2023fsc}. To further clarify the nature of $\Lambda_c(2910)^+$, more discussions are necessary.

\subsection{$1D$-wave $\lambda$-mode excitations}

According to the symmetry of wave functions, there are two $\lambda$-mode $1D$-wave $\Lambda_c$ states(see Table~\ref{table1}): $\Lambda_c|J^P=3/2^+,2\rangle_{\lambda\lambda}$ and $\Lambda_c|J^P=5/2^+,2\rangle_{\lambda\lambda}$. As shown in Table~\ref{table1}, the masses of the two $1D$ $\lambda$-mode $\Lambda_c$ states fluctuate around $\sim2.85$ GeV. Based on the predicted masses, $\Lambda_c|J^P=3/2^+,2\rangle_{\lambda\lambda}$ and $\Lambda_c|J^P=5/2^+,2\rangle_{\lambda\lambda}$ are probably assignments of observed states $\Lambda_c(2860)^+$ and $\Lambda_c(2880)^+$, respectively. Hence, we fix the masses of $\Lambda_c|J^P=3/2^+,2\rangle_{\lambda\lambda}$ and $\Lambda_c|J^P=5/2^+,2\rangle_{\lambda\lambda}$ at $M=2856$ MeV and $M=2882$ MeV, respectively, and list their decays in Table~\ref{1Dlambda}.

\begin{table}[]
	\caption{\label{1Dlambda}The partial decay widths of $\Lambda_c(2860)^+$ and $\Lambda_c(2880)^+$ assigned as $\lambda$-mode $1D$-wave $\Lambda_c$ states  $\Lambda_c|J^P=3/2^+,2\rangle_{\lambda\lambda}$ and $\Lambda_c|J^P=5/2^+,2\rangle_{\lambda\lambda}$, respectively. The unit is MeV.}
	\centering
	\begin{tabular}{c|ccccccc}	\hline\hline
\multirow{2}{*}{Decay width}&$\Lambda_c|J^P=3/2^+,2\rangle_{\lambda\lambda}$ &$\Lambda_c|J^P=5/2^+,2\rangle_{\lambda\lambda}$ \\ \cline{2-3}
                            &$\Lambda_c(2860)^+$&M=2882  \\ \hline
 $\Gamma[\Sigma_c\pi]$     &24.32&1.74  \\
$\Gamma[\Sigma_c^*\pi]$     &3.86&24.28 \\
$\Gamma[pD^0]$     &21.61&0.35 \\
$\Gamma[nD^+]$     &18.77&0.27 \\
$\Gamma_{\mathrm{Total}}$     &68.56&26.64  \\  \hline
Expt.&$67.6^{+10.1}_{-8.1}\pm1.4^{+5.9}_{-20.0}$&$5.6^{+0.8}_{-0.6}$\\
\hline\hline
\end{tabular}	
\end{table}

It is obtained that the total width of $\Lambda_c|J^P=3/2^+,2\rangle_{\lambda\lambda}$ is
\begin{equation}
\Gamma_{\mathrm{Total}}\simeq68.56~\mathrm{MeV},
\end{equation}
which is in good agree with the experimental central value.
Furthermore,  $pD^0$ ia one of the main decay mode, and the predicted branching fraction is
\begin{equation}
\frac{\Gamma[\Lambda_c|J^P=\frac{3}{2}^+,2\rangle_{\lambda\lambda}]\rightarrow pD^0}{\Gamma_{\mathrm{Total}}}\simeq32\%.
\end{equation}
This result is consistent with the fact that $\Lambda_c(2860)^+$ was observed in $pD^0$ invariant mass distribution~\cite{LHCb:2017jym}.
In addition, we get that
\begin{equation}
\frac{\Gamma[\Lambda_c|J^P=\frac{3}{2}^+,2\rangle_{\lambda\lambda}\rightarrow \Sigma_c\pi]}{\Gamma[\Lambda_c|J^P=\frac{3}{2}^+,2\rangle_{\lambda\lambda}\rightarrow pD^0]}\simeq1.13,
\end{equation}
\begin{equation}
\frac{\Gamma[\Lambda_c|J^P=\frac{3}{2}^+,2\rangle_{\lambda\lambda}\rightarrow nD^+]}{\Gamma[\Lambda_c|J^P=\frac{3}{2}^+,2\rangle_{\lambda\lambda}\rightarrow pD^0]}\simeq0.87.
\end{equation}
 If the observed state $\Lambda_c(2860)^+$ corresponds to $\Lambda_c|J^P=3/2^+,2\rangle_{\lambda\lambda}$ indeed, besides the $pD^0$ channel, the $\Sigma_c\pi$ and $nD^+$ may be another two interesting channels for observation of $\Lambda_c(2860)^+$ in future experiments. The $\Lambda_c(2860)^+$ resonance should be observed in the $\Lambda_c\pi\pi$ and $nD^+$ final states as well.

 For the state $\Lambda_c|J^P=5/2^+,2\rangle_{\lambda\lambda}$(see the Table~\ref{1Dlambda}), fixing its mass on M=2882 MeV, the total decay width
 \begin{equation}
\Gamma_{\mathrm{Total}}\simeq26.64~\mathrm{MeV},
\end{equation}
is about five times of the observation for $\Lambda_c(2880)^+$. Meanwhile, the predicted partial decay width ratio between $pD^0$ and $\Sigma_c\pi$ is
\begin{equation}
\frac{\Gamma[\Lambda_c|J^P=\frac{5}{2}^+,2\rangle_{\lambda\lambda}\rightarrow pD^0]}{\Gamma[\Lambda_c|J^P=\frac{5}{2}^+,2\rangle_{\lambda\lambda}\rightarrow \Sigma_c\pi]}\simeq0.20.
\end{equation}
This value is much smaller than the measured ratio($0.75\pm0.03\pm0.07$) by the Belle Collaboration~\cite{Belle:2024cmc}. Meanwhile, our theoretical calculation indicates that the $\Sigma_c^*\pi$ decay channel almost saturates the total decay widths. The partial decay width ratio between $\Sigma_c^*\pi$ and $\Sigma_c\pi$ is
\begin{equation}
\frac{\Gamma[\Lambda_c|J^P=\frac{5}{2}^+,2\rangle_{\lambda\lambda}\rightarrow \Sigma_c^*\pi]}{\Gamma[\Lambda_c|J^P=\frac{5}{2}^+,2\rangle_{\lambda\lambda}\rightarrow \Sigma_c\pi]}\simeq13.95,
\end{equation}
which is inconsistent with the analysis from the CLEO Collaboration~\cite{CLEO:2000mbh}. Hence, according to our investigation, the experimental widths and some partial decay width ratios cannot be reproduced. To further clarify the properties of the $\Lambda_c(2880)^+$ resonance, more experimental and theoretical investigations may be necessary.

\subsection{$1D$-wave $\rho$-mode excitations}

Within the quark model, there are two $\rho$-mode $1D$-wave $\Lambda_c$ states: $\Lambda_c|J^P=3/2^+,2\rangle_{\rho\rho}$ and $\Lambda_c|J^P=5/2^+,2\rangle_{\rho\rho}$. According to the mass predictions listed in Table~\ref{table1}, their masses are about $M\sim3.10$ GeV. Firstly, we fix the masses at the predictions within a relativized quark potential model in Ref.~\cite{Capstick:1986ter}, and collect the decay properties in Table~\ref{1Drho}.

\begin{table}[h]
	\caption{\label{1Drho}The partial decay widths of the two $\rho$-mode $1D$-wave $\Lambda_c$ states, which masses are taken from the predictions in Ref.~\cite{Capstick:1986ter}. $\Gamma_{\text{Total}}$ represents the total decay width and the unit is MeV.}
	\centering
	\begin{tabular}{c|ccccccc}	\hline\hline
\multirow{2}{*}{Decay width}&$\Lambda_c|J^P=3/2^+,2\rangle_{\rho\rho}$ &$\Lambda_c|J^P=5/2^+,2\rangle_{\rho\rho}$ \\ \cline{2-3}
                            &M=3035&M=3140  \\ \hline
 $\Gamma[\Sigma_c\pi]$     &82.20&33.38  \\
$\Gamma[\Sigma_c^*\pi]$     &25.66&107.24 \\
$\Gamma[\Lambda_c\omega]$     &-&31.25 \\
$\Gamma[\Xi_c'K]$     &-&0.02 \\
$\Gamma[\Sigma_c|J^P=1/2^-,0\rangle_\lambda\pi]$     &0.01&0.29 \\
$\Gamma[\Sigma_c|J^P=1/2^-,1\rangle_\lambda\pi]$     &0.13&0.79 \\
$\Gamma[\Sigma_c|J^P=3/2^-,1\rangle_\lambda\pi]$     &0.06&1.91 \\
$\Gamma[\Sigma_c|J^P=3/2^-,2\rangle_\lambda\pi]$     &11.81&0.11 \\
$\Gamma[\Sigma_c|J^P=5/2^-,2\rangle_\lambda\pi]$     &0.00&2085 \\
$\Gamma_{\mathrm{Total}}$     &119.87&195.84  \\
\hline\hline
\end{tabular}	
\end{table}

The $\Lambda_c|J^P=3/2^+,2\rangle_{\rho\rho}$ state may be a moderate state with a width of $\Gamma_{\mathrm{Total}}\simeq120$ MeV, and mainly decays into $\Sigma_c\pi$. The predicted branching fraction is
\begin{equation}
\frac{\Gamma[\Lambda_c|J^P=\frac{3}{2}^+,2\rangle_{\rho\rho}\rightarrow \Sigma_c\pi]}{\Gamma_{\mathrm{Total}}}\simeq69\%.
\end{equation}
Hence, the $\Lambda_c|J^P=3/2^+,2\rangle_{\rho\rho}$ state are likely to be observed in the $\Lambda_c\pi\pi$ final state via the decay chain $\Lambda_c|J^P=3/2^+,2\rangle_{\rho\rho}\rightarrow \Sigma_c\pi\rightarrow\Lambda_c\pi\pi$.

Meanwhile, the partial decay width of $\Sigma_c^*\pi$ is sizable, and the branching fraction is
\begin{equation}
\frac{\Gamma[\Lambda_c|J^P=\frac{3}{2}^+,2\rangle_{\rho\rho}\rightarrow \Sigma_c^*\pi]}{\Gamma_{\mathrm{Total}}}\simeq21\%.
\end{equation}
Thus, $\Lambda_c|J^P=3/2^+,2\rangle_{\rho\rho}\rightarrow \Sigma_c^*\pi\rightarrow\Lambda_c\pi\pi$ may be another interesting decay chain for experimental exploration.

The other $\rho$-mode $1D$-wave state $\Lambda_c|J^P=5/2^+,2\rangle_{\rho\rho}$ has a width of $\Gamma_{\mathrm{Total}}\simeq196$ MeV, and mainly decays into $\Sigma_c^*\pi$ with a branching fraction
\begin{equation}
\frac{\Gamma[\Lambda_c|J^P=\frac{5}{2}^+,2\rangle_{\rho\rho}\rightarrow \Sigma_c^*\pi]}{\Gamma_{\mathrm{Total}}}\simeq55\%.
\end{equation}

Furthermore, $\Lambda_c|J^P=5/2^+,2\rangle_{\rho\rho}$ may have a considerable decay rate into $\Sigma_c\pi$ and $\Lambda_c\omega$. The predicted branching fractions are
\begin{equation}
\frac{\Gamma[\Lambda_c|J^P=\frac{5}{2}^+,2\rangle_{\rho\rho}\rightarrow \Sigma_c\pi/\Lambda_c\omega]}{\Gamma_{\mathrm{Total}}}\simeq17/16\%.
\end{equation}
However, this state may be too broad to be observed in experiments.

\begin{figure}[h]
	\centering \epsfxsize=8.5 cm \epsfbox{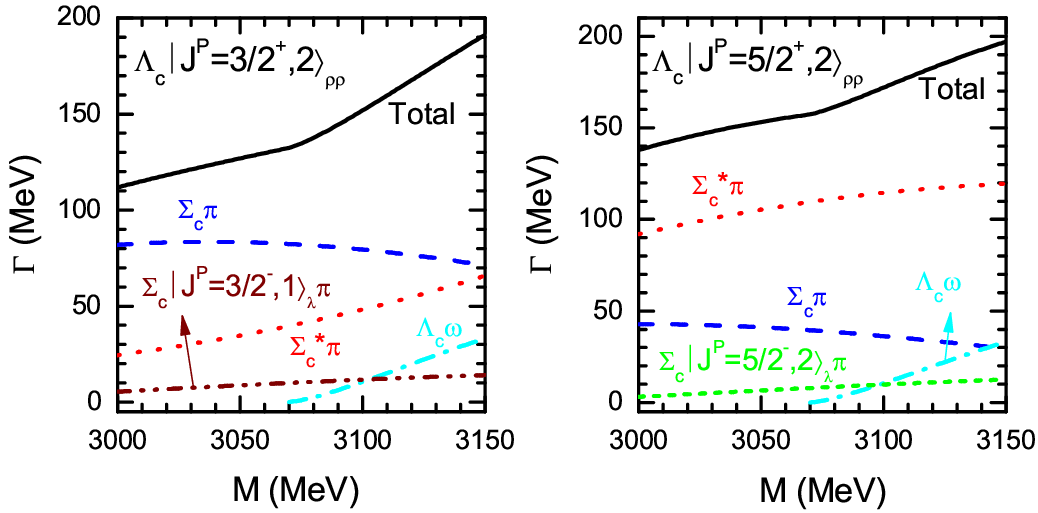}
	\caption{ Partial and total strong decay widths of the two $\rho$-mode $1D$-wave $\Lambda_c$ states as functions of the masses. Some decay channels are too small to show in figure.}
    \label{FIG3}
\end{figure}

Then accounting for the uncertainty of the predicted masses, which may bring uncertainties to the theoretical results, we plot the decay properties of the $1D$-wave $\rho$-mode $\Lambda_c$ as functions of masses within the range of $M=(3.00-3.15)$ GeV, as shown in Fig.~\ref{FIG3}. We can find that the total decay widths vary within the scope of $\Gamma_{\mathrm{Total}}<200$ MeV as the mass increasing. In addition, when the masses of the $1D$-wave $\rho$-mode $\Lambda_c$ states are above the threshold of $\Lambda_c\omega$, the partial decay width of this channel will be sizable and increases dramatically with the mass.

\subsection{$2S$-wave excitations}
The $2S$-wave states are the first radial excited states, and the radial quantum number $n_{\lambda}=1$ or $n_{\rho}=1$. Hence, according to the symmetry of the wave function, there are two $2S$-wave $\Lambda_c$ states: $\Lambda_{c1}|J^P=1/2^+,0\rangle_{\lambda}$(denote $n_{\lambda}=1$) and $\Lambda_{c1}|J^P=1/2^+,0\rangle_{\rho}$(denote $n_{\rho}=1$). As shown in  Table~\ref{table1}, the mass of the $\Lambda_{c1}|J^P=1/2^+,0\rangle_{\lambda}$ state is around $M\sim(2.76\sim2.79)$ MeV, while the mass of the $\Lambda_{c1}|J^P=1/2^+,0\rangle_{\rho}$ state is slightly heavier and about $M\sim2.97$ MeV.
Fixing the masses at the predictions in Ref.~\cite{Capstick:1986ter}, we analyze the decay properties of the two $2S$-wave $\Lambda_c$ states, and collected their partial strong decay widths in Table~\ref{2s}.

The $\Lambda_{c1}|J^P=1/2^+,0\rangle_{\lambda}$ state probably has a narrow width of several tens of MeV, and mainly decays into the $\Sigma_c\pi$ and $\Sigma_c^*\pi$ channels. In this case, this state has a good potential to be observed in the $\Lambda_c\pi\pi$ final state by the intermediate channels $\Sigma_c\pi$ and $\Sigma_c^*\pi$. Combining the predicted mass and our calculations, the $\Lambda_{c1}|J^P=1/2^+,0\rangle_{\lambda}$ state may be a assignment of the observed state $\Lambda_c(2765)^+$. Hence, we further take this state as $\Lambda_c(2765)^+$, and list its decay properties in Table~\ref{2s} as well. With the mass of $\Lambda_{c1}|J^P=1/2^+,0\rangle_{\lambda}$ fixed on $M=2767$ MeV, the total decay width
\begin{equation}
\Gamma_{\mathrm{Total}}\simeq21.14 \mathrm{MeV}
\end{equation}
is about half of the experimental value ($\Gamma_{\mathrm{Expt.}}=50$ MeV). The predicted partial decay width ratio between the dominant modes $\Sigma_c\pi$ and $\Sigma_c^*\pi$ is
\begin{equation}
\frac{\Gamma[\Lambda_{c1}|J^P=1/2^+,0\rangle_{\lambda}\rightarrow \Sigma_c\pi]}{\Gamma[\Lambda_{c1}|J^P=1/2^+,0\rangle_{\lambda}\rightarrow \Sigma_c^*\pi]}\simeq1.01.
\end{equation}
This result is consistent with the fact that $\Lambda_c(2765)^+$ is firstly observed in the $\Sigma_c\pi$ and $\Sigma_c^*\pi$ channels by the CLEO Collaboration~\cite{CLEO:2000mbh}.
Thus, the state $\Lambda_{c1}|J^P=1/2^+,0\rangle_{\lambda}$ may be a candidate of $\Lambda_c(2765)^+$.

\begin{table}[h]
	\caption{\label{2s}The partial decay widths of the two $2S$-wave $\Lambda_c$ states, which masses are taken from the predictions in Ref.~\cite{Capstick:1986ter}. $\Gamma_{\text{Total}}$ represents the total decay width and the unit is MeV.}
	\centering
	\begin{tabular}{c|cc|ccccc}	\hline\hline
\multirow{2}{*}{Decay width}&\multicolumn{2}{c}{$\Lambda_{c1}|J^P=1/2^+,0\rangle_{\lambda}$} &$\Lambda_{c1}|J^P=1/2^+,0\rangle_{\rho}$ \\ \cline{2-4}
                            &M=2775&$\Lambda_c(2765)^+$&M=2970  \\ \hline
 $\Gamma[\Sigma_c\pi]$     &11.63&10.90&81.48  \\
$\Gamma[\Sigma_c^*\pi]$     &11.57&10.24&149.88 \\
$\Gamma[\Sigma_c|J^P=1/2^-,0\rangle_\lambda\pi]$     &-&-&13.15 \\
$\Gamma[\Sigma_c|J^P=1/2^-,1\rangle_\lambda\pi]$     &-&-&0.00 \\
$\Gamma[\Sigma_c|J^P=3/2^-,1\rangle_\lambda\pi]$     &-&-&0.00 \\
$\Gamma[\Sigma_c|J^P=3/2^-,2\rangle_\lambda\pi]$     &-&-&0.01 \\
$\Gamma_{\mathrm{Total}}$     &23.20&21.14&244.52  \\ \hline
Expt.   &\multicolumn{2}{c|}{50} &- \\
\hline\hline
\end{tabular}	
\end{table}

For the other $2S$-wave state $\Lambda_{c1}|J^P=1/2^+,0\rangle_{\rho}$, the main decay channels are $\Sigma_c\pi$ and $\Sigma_c^*\pi$ as well, and the partial widths ratio is
\begin{equation}
\frac{\Gamma[\Lambda_{c1}|J^P=1/2^+,0\rangle_{\rho}\rightarrow \Sigma_c\pi]}{\Gamma[\Lambda_{c1}|J^P=1/2^+,0\rangle_{\rho}\rightarrow \Sigma_c^*\pi]}\simeq0.54.
\end{equation}
While, this state is most likely to be a broad state with a width of about $\Gamma_{\mathrm{Total}}\simeq245$ MeV. Thus, it is hard to observe the $\Lambda_{c1}|J^P=1/2^+,0\rangle_{\rho}$ state in experiment for its broad decay width.

\begin{figure}[h]
	\centering \epsfxsize=8.5 cm \epsfbox{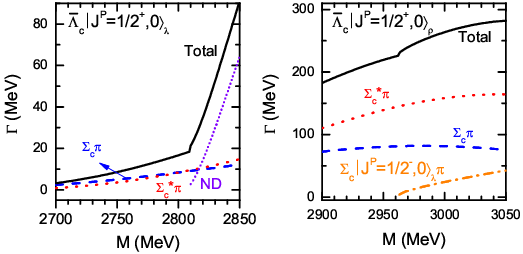}
	\caption{ Partial and total strong decay widths of the two $2S$-wave $\Lambda_c$ states as functions of the masses. Some decay channels are too small to show in figure.}
    \label{FIG4}
\end{figure}

Considering the uncertainty of the masses of the $2S$-wave $\Lambda_c$ states, we further investigate the strong decay widths as a function of the mass in Fig.~\ref{FIG4}. It is shown that the decay properties of the $2S$-wave $\Lambda_c$ excitations are sensitive to their masses varying in the considered region. Furthermore, if the mass of $\Lambda_{c1}|J^P=1/2^+,0\rangle_{\lambda}$ is above the threshold of $ND$, the corresponding partial decay width of $ND$ will increase dramatically with the mass and holds a important place in the strong decay.

\subsection{$2P$-wave $\lambda$-mode excitations}
In this calculations, the $2P$-wave $\lambda$-mode $\Lambda_c$ excitations correspond to the radial quantum number $n_{\lambda}=1$ and orbital quantum number $l_{\lambda}=1$. Hence, based on the quark model classification there are two $2P$-wave $\lambda$-mode $\Lambda_c$ states: $\Lambda_{c1}|J^P=1/2^-,1\rangle_{\lambda}$ and $\Lambda_{c1}|J^P=3/2^-,1\rangle_{\lambda}$. Their theoretical masses and possible two-body decay channels are listed in Table~\ref{table1}.

\begin{figure}[h]
	\centering \epsfxsize=8.5 cm \epsfbox{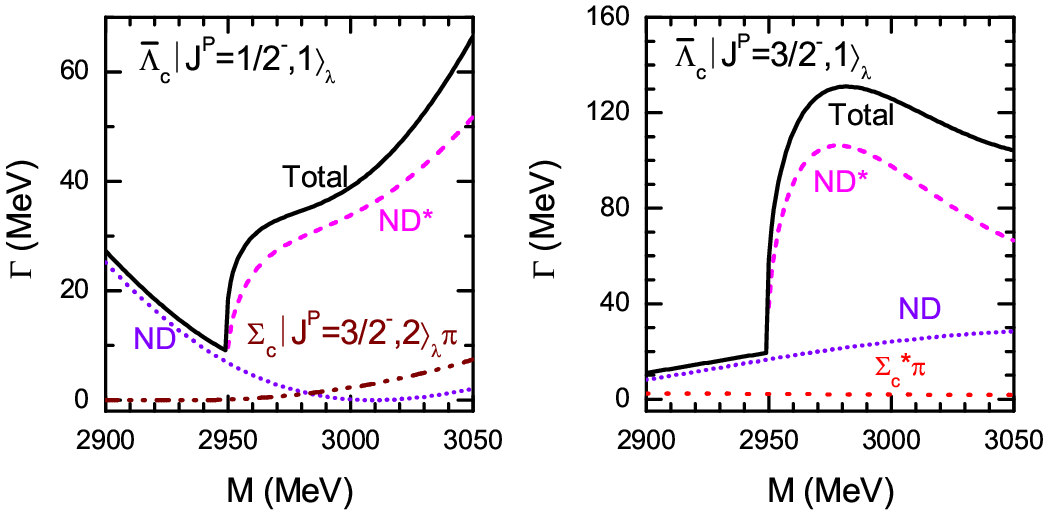}
	\caption{ Partial and total strong decay widths of the two $\lambda$-mode $2P$-wave $\Lambda_c$ states as functions of the masses. Some decay channels are too small to show in figure.}
    \label{FIG5}
\end{figure}

From the table, it is known that the predicted masses of the two $2P$-wave $\lambda$-mode $\Lambda_c$ baryons are about $M\sim3.00$ GeV, which is close to the measured mass of the observed state $\Lambda_c(2940)^+$. As the possible assignment, it is crucial to investigate the decay behaviors of the two $2P$-wave $\lambda$-mode $\Lambda_c$ baryons. Considering the uncertainties of the predicted masses of $\Lambda_{c1}|J^P=1/2^-,1\rangle_{\lambda}$ and $\Lambda_{c1}|J^P=3/2^-,1\rangle_{\lambda}$, we plot the decay width as a function of the mass in the range of $M=(2.90-3.05)$ GeV in Fig.~\ref{FIG5}.

From the Figure, the total decay widths of the two $2P$-wave $\lambda$-mode $\Lambda_c$ baryons are about dozens of MeV within the mass range what we considered. When their masses are lie below the threshold of $ND^*$, their strong decays are dominated by the $ND$ channel. However, if their masses are above the threshold of $ND^*$, the dominant decay channel should be $ND^*$, and their total decay widths are strong dependence on the masses.

In addition, we notice that the measured mass and total decay width of $\Lambda_c(2940)^+$ are consistent with the properties the two $2P$-wave $\lambda$-mode $\Lambda_c$ baryons. Hence, we further fix the masses of $\Lambda_{c1}|J^P=1/2^-,1\rangle_{\lambda}$ and $\Lambda_{c1}|J^P=3/2^-,1\rangle_{\lambda}$ at $M=2940$ MeV, and collect their decay properties in Table~\ref{2p}.

\begin{table}[h]
	\caption{\label{2p}The partial decay widths of $\Lambda_c(2940)^+$ assigned as the two $2P$-wave $\lambda$-mode $\Lambda_c$ states $\Lambda_{c1}|J^P=1/2^-,1\rangle_{\lambda}$ and $\Lambda_{c1}|J^P=3/2^-,1\rangle_{\lambda}$, respectively.}
	\centering
	\begin{tabular}{c|ccccccc}	\hline\hline
\multirow{2}{*}{Decay width}&$\Lambda_{c1}|J^P=1/2^-,1\rangle_{\lambda}$ &$\Lambda_{c1}|J^P=3/2^-,1\rangle_{\lambda}$ \\ \cline{2-3}
                           &M=2940&$\Lambda_c(2940)^+$  \\ \hline
 $\Gamma[\Sigma_c\pi]$     &0.68&2.28  \\
$\Gamma[\Sigma_c^*\pi]$     &3.18&1.61 \\
$\Gamma[pD^0]$     &5.72&10.21 \\
$\Gamma[nD^+]$     &6.89&9.55 \\
$\Gamma_{\mathrm{Total}}$     &16.47&23.65  \\ \hline
Expt.   &\multicolumn{2}{c}{$20^{+6}_{-5}$} \\
\hline\hline
\end{tabular}	
\end{table}

It is seen that the total decay width of the $\Lambda_{c1}|J^P=1/2^-,1\rangle_{\lambda}$ is
\begin{equation}
\Gamma_{\mathrm{Total}}\simeq16.47~\mathrm{MeV},
\end{equation}
 which is in agreement with the experimental observation. The dominant decay channels are $pD^0$ and $nD^+$ with predicted branching fractions are
 \begin{equation}
\frac{\Gamma[\Lambda_{c1}|J^P=1/2^-,1\rangle_{\lambda}\rightarrow pD^0/nD^+]}{\Gamma_{\mathrm{Total}}}\simeq35/42\%.
\end{equation}
Meanwhile, the $\Sigma_c^*\pi$ decay channel occupies a sizable branching fraction, and the corresponding branching fraction is about
 \begin{equation}
\frac{\Gamma[\Lambda_{c1}|J^P=1/2^-,1\rangle_{\lambda}\rightarrow \Sigma_c^*\pi]}{\Gamma_{\mathrm{Total}}}\simeq19\%.
\end{equation}

However, we notice that the partial decay width of $\Lambda_{c1}|J^P=1/2^-,1\rangle_{\lambda}\rightarrow \Sigma_c\pi$ seems a little small, which does not accord with the fact that $\Lambda_c(2940)$ was observed in the $\Sigma_c\pi$ channel. Meanwhile, the partial decay width ratio between $pD^0$ and $\Sigma_c\pi$
\begin{equation}
\frac{\Gamma[\Lambda_{c1}|J^P=1/2^-,1\rangle_{\lambda}\rightarrow pD^0]}{\Gamma[\Lambda_{c1}|J^P=1/2^-,1\rangle_{\lambda}\rightarrow \Sigma_c\pi]}\simeq8.41
\end{equation}
is significantly greater than the latest experimental value ($3.59\pm0.21\pm0.56$) measured by the LHCb Collaboration~\cite{Belle:2024cmc}.

The total decay width of $\Lambda_{c1}|J^P=3/2^-,1\rangle_{\lambda}$
\begin{equation}
\Gamma_{\mathrm{Total}}\simeq23.65~\mathrm{MeV}
\end{equation}
agrees with the experimental value as well. Meanwhile, the decays are governed by $pD^0$ and $nD^+$ and the predicted branching fractions are
 \begin{equation}
\frac{\Gamma[\Lambda_{c1}|J^P=3/2^-,1\rangle_{\lambda}\rightarrow pD^0/nD^+]}{\Gamma_{\mathrm{Total}}}\simeq43/40\%.
\end{equation}
In addition,  $\Lambda_{c1}|J^P=3/2^-,1\rangle_{\lambda}$ has a sizable decay width into $\Sigma_c\pi$. The predicted partial decay width ratio between $pD^0$ and $\Sigma_c\pi$ is
\begin{equation}
\frac{\Gamma[\Lambda_{c1}|J^P=3/2^-,1\rangle_{\lambda}\rightarrow pD^0]}{\Gamma[\Lambda_{c1}|J^P=3/2^-,1\rangle_{\lambda}\rightarrow \Sigma_c\pi]}\simeq4.48,
\end{equation}
which is close to the upper limit of the measurement~\cite{Belle:2024cmc}.

In conclusion, our calculation indicates that the strong decay properties of $\Lambda_{c1}|J^P=3/2^-,1\rangle_{\lambda}$ is in good agreement with the nature of $\Lambda_c(2940)$, and  $\Lambda_{c1}|J^P=3/2^-,1\rangle_{\lambda}$ can be a good candidate. It should be pointed out that the threshold of the main decay channel $DN$ is close to the mass of $\Lambda_c(2940)$, which may suggest the importance of the coupled-channel effects for understanding the $\Lambda_c(2940)^+$ state.

\subsection{$2P$-wave $\rho$-mode excitations}

In the present work, the $2P$-wave $\rho$-mode excitations correspond to the radial quantum number $n_{\rho}$=1 and orbital quantum number $l_{\rho}$=1. According to the quark model, there are five $2P$-wave $\rho$-mode $\Lambda_c$ baryons: $\Lambda_{c1}|J^P=1/2^-,0\rangle_{\rho}$, $\Lambda_{c1}|J^P=1/2^-,1\rangle_{\rho}$, $\Lambda_{c1}|J^P=3/2^-,1\rangle_{\rho}$, $\Lambda_{c1}|J^P=3/2^-,2\rangle_{\rho}$ and $\Lambda_{c1}|J^P=5/2^-,2\rangle_{\rho}$. For their masses, there are a few discussions in theoretical references and we have collected in Table~\ref{table1} as well. From the table, the masses of the $2P$-wave $\rho$-mode $\Lambda_c$ excitations are about $M\sim3.20$ GeV. Fixing the masses of the $2P$-wave $\rho$-mode $\Lambda_c$ excitations on the predicted masses from Ref.~\cite{Capstick:1986ter}, we discuss their decay properties and list the results in Table~\ref{Prho}.

For the $\Lambda_{c1}|J^P=1/2^-,0\rangle_{\rho}$ state, the total decay width is about $\Gamma_{\mathrm{Total}}\simeq$175 MeV. The dominant decay modes are $\Sigma_c|J^P=1/2^-,1\rangle_{\rho}\pi$ and $\Sigma_c|J^P=3/2^-,1\rangle_{\rho}\pi$ with the partial decay ratio
\begin{equation}
\frac{\Gamma[\Lambda_{c1}|J^P=1/2^-,0\rangle_{\rho}\rightarrow\Sigma_c|J^P=1/2^-,1\rangle_{\rho}\pi]}{\Gamma[\Lambda_{c1}|J^P=1/2^-,0\rangle_{\rho}\rightarrow\Sigma_c|J^P=3/2^-,1\rangle_{\rho}\pi]}\simeq0.51.
\end{equation}
Hence, this state may be observed in the $\Lambda_c\pi\pi\pi$ final state via the decay chains $\Lambda_{c1}|J^P=1/2^-,0\rangle_{\rho}\rightarrow\Sigma_c|J^P=1/2^-,1\rangle_{\rho}\pi\rightarrow \Sigma_c\pi\pi\rightarrow \Lambda_c\pi\pi\pi$ and $\Lambda_{c1}|J^P=1/2^-,0\rangle_{\rho}\rightarrow\Sigma_c|J^P=3/2^-,1\rangle_{\rho}\pi\rightarrow \Sigma_c^*\pi\pi\rightarrow \Lambda_c\pi\pi\pi$.

Meanwhile, the partial decay width of $\Gamma[\Lambda_{c1}|J^P=1/2^-,0\rangle_{\rho}\rightarrow \Lambda_c\eta]$ is considerable. The branching fraction is
\begin{equation}
\frac{\Gamma[\Lambda_{c1}|J^P=1/2^-,0\rangle_{\rho}\rightarrow \Lambda_c\eta]}{\Gamma_{\mathrm{Total}}}\simeq2\%.
\end{equation}
The $\Lambda_c\eta$ channel may be also notable decay mode for future exploring the $\bar{\Lambda}_c|J^P=1/2^-,0\rangle_{\rho}$ state.

The states $\Lambda_{c1}|J^P=1/2^-,1\rangle_{\rho}$ and $\Lambda_{c1}|J^P=3/2^-,1\rangle_{\rho}$ are most likely to be the moderate states with a total decay width of $\Gamma_{\mathrm{Total}}\sim70$ MeV. While, their dominant decay channels are different. $\Lambda_{c1}|J^P=1/2^-,1\rangle_{\rho}$ mainly decays via the $\Sigma_c\pi$ and $\Sigma_c^*\pi$ channels, and the branching fractions are
\begin{equation}
\frac{\Gamma[\Lambda_{c1}|J^P=1/2^-,1\rangle_{\rho}\rightarrow\Sigma_c\pi/\Sigma_c^*\pi]}{\Gamma_{\mathrm{Total}}}\simeq26/22\%.
\end{equation}
In addition, the $\Lambda_{c1}|J^P=1/2^-,1\rangle_{\rho}$ state has sizable partial widths decaying into $\Sigma_c|J^P=1/2^-,1\rangle_{\rho}\pi$, $\Lambda_c\omega$ and $\Sigma_c|J^P=1/2^-,1\rangle_{\lambda}\pi$. Those channels' predicted branching fractions are about $(9\sim15)$\%.

As to $\Lambda_{c1}|J^P=3/2^-,1\rangle_{\rho}$, it decays mainly through the $\Sigma_c^*\pi$ channel. The predicted branching fraction is
\begin{equation}
\frac{\Gamma[\Lambda_{c1}|J^P=3/2^-,1\rangle_{\rho}\rightarrow \Sigma_c^*\pi]}{\Gamma[\mathrm{Total}]}\simeq30\%.
\end{equation}
Meanwhile, the partial decay widths of $\Sigma_c|J^P=3/2^-,1\rangle_{\rho}\pi$, $\Sigma_c|J^P=3/2^-,1\rangle_{\lambda}\pi$, $\Lambda_c\omega$ and $\Sigma_c\pi$ are considerable, and the corresponding branching fractions are about $(10\sim20)$\%.

\begin{figure}[]
	\centering \epsfxsize=8.5 cm \epsfbox{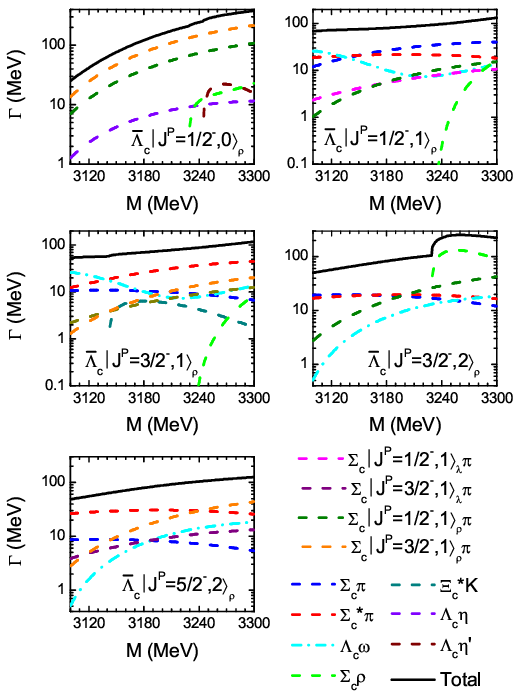}
	\caption{ Partial and total strong decay widths of the $\rho$-mode $2P$-wave $\Lambda_c$ states as functions of the masses. Some decay channels are too small to show in figure.}
    \label{FIG6}
\end{figure}

The decay width of the state $\Lambda_{c1}|J^P=3/2^-,2\rangle_{\rho}$ is about $\Gamma\simeq293$ MeV. Its strong decays are governed by the $\Sigma_c\rho$ channel with the branching fraction around $\sim60\%$. Meanwhile, this state has a sizable decay rate decaying into the $\Sigma_c|J^P=1/2^-,1\rangle_{\rho}\pi$ channel, and the predicted branching ratio is about $\sim12\%$. However, the $\Lambda_{c1}|J^P=3/2^-,2\rangle_{\rho}$ might be too broad to observed in experiments.

The state $\Lambda_{c1}|J^P=5/2^-,2\rangle_{\rho}$ may be a narrow state with a total decay width around $\Gamma\simeq52$ MeV, and mainly decays into the $\Sigma_c^*\pi$ channel. The predicted branching fraction is
\begin{equation}
\frac{\Gamma[\Lambda_{c1}|J^P=5/2^-,2\rangle_{\rho}\rightarrow \Sigma_c^*\pi]}{\Gamma[\mathrm{Total}]}\simeq43\%.
\end{equation}
Thus, this state is most likely to be observed in the $\Lambda_c\pi\pi$ final state via the decay chain $\Lambda_{c1}|J^P=5/2^-,2\rangle_{\rho}\rightarrow \Sigma_c^*\pi\rightarrow \Lambda_c\pi\pi$. In addition, the partial widths of $\Sigma_c\pi$ and $\Sigma_c|J^P=3/2^-,1\rangle_{\lambda/\rho}\pi$ are sizable as well, and all of the branching fractions are about $13\%$.
Hence, the $\Lambda_c\pi\pi$ and $\Lambda_c\pi\pi\pi$ final states via the decay chains $\Lambda_{c1}|J^P=5/2^-,2\rangle_{\rho}\rightarrow \Sigma_c\pi\rightarrow \Lambda_c\pi\pi$ and $\Lambda_{c1}|J^P=5/2^-,2\rangle_{\rho}\rightarrow \Sigma_c|J^P=3/2^-,1\rangle_{\lambda/\rho}\pi\rightarrow \Sigma_c^*\pi\pi\rightarrow \Lambda_c\pi\pi\pi$ may be another three interesting decay channels for experimental observations.

Similarly, we also plot the partial decay widths of the $2P$-wave $\rho$-mode $\Lambda_c$ baryons as a function of the mass in region of $M=(3.10-3.30)$ GeV. The sensitivities of the decay properties if these states to their masses are shown in Fig.~\ref{FIG6}. From the figure, the partial widths of dominant decay channels for most of the states vary gently with mass increasing. In addition, if the $2P$-wave $\rho$-mode $\Lambda_c$ baryons are above the threshold of $\Sigma_c\rho$, most of the states can decay via $\Sigma_c\rho$ with a partial width about several MeV or more.

\section{Summary}

In the present work, we have systematically studied the strong decay properties of the low-lying $1P$-, $1D$-, $2S$- and $2P$-wave $\Lambda_c$ baryons in the framework of the quark pair creation model within the $j-j$ coupling scheme. Our main results are summarized as follows.

For the two $1P$-wave $\lambda$-mode $\Lambda_c$ baryons $\Lambda_c|J^P=1/2^-,1\rangle_{\lambda}$ and $\Lambda_c|J^P=3/2^-,1\rangle_{\lambda}$, they are corresponding to the well determined states $\Lambda_c(2595)^+$ and $\Lambda_c(2625)^+$, respectively, and we can reproduce the experimental data well in theory. Meanwhile, we notice that the mass of $\Lambda_c(2595)^+$ is very close to the threshold of $\Sigma_c\pi$ and this causes the decay widths are highly sensitive to its mass precision.

For the $1P$-wave $\rho$-mode $\Lambda_c$ baryons, $\Lambda_c|J^P=1/2^-,0\rangle_{\rho}$ is most likely to be a very narrow state, and the $\Sigma_c\pi$ decay mode almost saturates its total decay widths. Hence, the $\Lambda_c\pi\pi$ final state may be a ideal decay channel to explore $\Lambda_c|J^P=1/2^-,0\rangle_{\rho}$ in future experiments. The states $\Lambda_c|J^P=1/2^-,1\rangle_{\rho}$ and $\Lambda_c|J^P=3/2^-,1\rangle_{\rho}$ are probably two quite broad states with a width of around $\Gamma_{\mathrm{Total}}\sim780$ MeV. Their dominant decay channels are $\Sigma_c\pi$ and $\Sigma_c^*\pi$, respectively. Considering the decay widths too broad, the two states may be difficult to be observed in experiments. The total decay widths of $\Lambda_c|J^P=3/2^-,2\rangle_{\rho}$ and $\Lambda_c|J^P=5/2^-,2\rangle_{\rho}$ are several tens MeV, and their strong decays are governed by the $\Sigma_c\pi$ and $\Sigma_c^*\pi$ channels. Combining the mass and the decay properties of the newly observed state $\Lambda_c(2910)^+$, both $\Lambda_c|J^P=3/2^-,2\rangle_{\rho}$ and $\Lambda_c|J^P=5/2^-,2\rangle_{\rho}$ may be good candidates. Further determine which one of the two is it, the partial decay width ratio between $\Sigma_c\pi$ and $\Sigma_c^*\pi$ may be a good criterion.

As to the $1D$-wave $\lambda$-mode $\Lambda_c$ excitations, $\Lambda_c|J^P=3/2^+,2\rangle_{\lambda\lambda}$ is probably a good assignment of the observed state $\Lambda_c(2860)^+$. In addition, if the observed state $\Lambda_c(2860)^+$ corresponds to $\Sigma_c|J^P=3/2^+,2\rangle_{\lambda\lambda}$ indeed, besides the $pD^0$ channel, the $\Sigma_c\pi$ and $nD^+$ may be another two interesting channels for future experimental observation. The state $\Lambda_c|J^P=5/2^+,2\rangle_{\lambda\lambda}$ may be a moderate state with a width of about $(20\sim30)$ MeV, and mainly decays via the $\Sigma_c^*\pi$ channel. According to our investigation if we take the observed state $\Lambda_c(2880)^+$ as $\Lambda_c|J^P=5/2^+,2\rangle_{\lambda\lambda}$, the measured total decay widths and some partial decay width ratios cannot be reproduced. To further clarify the inner structure of the $\Lambda_c(2880)^+$ resonance, more experimental and theoretical efforts are needed.

For the $1D$-wave $\rho$-mode $\Lambda_c$ excitations, $\Lambda_c|J^P=3/2^+,2\rangle_{\rho\rho}$ may be a moderate state with a width of $\Gamma_{\mathrm{Total}}\sim120$ MeV, and mainly decays into $\Sigma_c\pi$. Meanwhile, the partial decay width of $\Sigma_c^*\pi$ is sizable. Hence, the  $\Lambda_c|J^P=3/2^+,2\rangle_{\rho\rho}$ has the possibility to be observed in the $\Lambda_c\pi\pi$ final state via the decay chains $\Sigma_c|J^P=3/2^+,2\rangle_{\rho\rho}\rightarrow \Sigma_c^{(*)}\pi\rightarrow\Lambda_c\pi\pi$. The other $1D$-wave $\rho$-mode state $\Lambda_c|J^P=5/2^+,2\rangle_{\rho\rho}$ has a width of $\Gamma_{\mathrm{Total}}\sim196$ MeV, and dominantly decays into $\Sigma_c^*\pi$. Moreover, the decay rates into $\Sigma_c\pi$ and $\Lambda_c\omega$ are considerable. However, this state may be too broad to be observed in experiments.

The $2S$-wave state $\Lambda_{c1}|J^P=1/2^+,0\rangle_{\lambda}$ probably has a width of dozens of MeV, and mainly decays into the $\Sigma_c\pi$ and  $\Sigma_c^*\pi$ channels. Combining the predicted mass and our calculations, the possibility of $\Lambda_{c1}|J^P=1/2^+,0\rangle_{\lambda}$ being a assignment of the observed state $\Lambda_c(2765)^+$ can't be excluded entirely.
For the other $2S$-wave state $\Lambda_{c1}|J^P=1/2^+,0\rangle_{\rho}$, the main decay channels are $\Sigma_c\pi$ and $\Sigma_c^*\pi$ as well. While this state is most likely to be a broad state with a width of about $\Gamma_{\mathrm{Total}}\sim245$ MeV. Thus, it is a great challenge to observed the $\Lambda_{c1}|J^P=1/2^+,0\rangle_{\rho}$ state in experiments for its broad decay width.

The total decay widths of the two $2P$-wave $\lambda$-mode $\Lambda_c$ states $\Lambda_{c1}|J^P=1/2^-,1\rangle_{\lambda}$ and $\Lambda_{c1}|J^P=3/2^-,1\rangle_{\lambda}$ are about dozens of MeV, and the $ND$ decay channel almost saturates their decay widths. Comparing the masses and total decay widths, $\Lambda_{c1}|J^P=1/2^-,1\rangle_{\lambda}$ and $\Lambda_{c1}|J^P=3/2^-,1\rangle_{\lambda}$ are allowed good assignments of $\Lambda_c(2940)^+$. While, the partial decay width ratio between $pD^0$ and $\Sigma_c\pi$ for $\Lambda_{c1}|J^P=3/2^-,1\rangle_{\lambda}$ is predicted to be $4.48$, which is close to the upper limit of the newest measured value by Belle Collaboration. In this case,  $\Lambda_{c1}|J^P=3/2^-,1\rangle_{\lambda}$ is more favorable.

For the $2P$-wave $\rho$-mode $\Lambda_c$ states, $\Lambda_{c1}|J^P=1/2^-,0\rangle_{\rho}$ probably has a width of $\Gamma_{\mathrm{Total}}\sim175$ MeV, and dominantly decays into $\Sigma_c|J^P=1/2^-,1\rangle_{\rho}\pi$ and $\Sigma_c|J^P=3/2^-,1\rangle_{\rho}\pi$. The states $\Lambda_{c1}|J^P=1/2^-,1\rangle_{\rho}$ and $\Lambda_{c1}|J^P=3/2^-,1\rangle_{\rho}$ are most likely to be moderate states with a width of about $\Gamma_{\mathrm{Total}}\sim70$ MeV. Except the main decay channels $\Sigma_c\pi$ and $\Sigma_c^*\pi$, the partial decay widths of the final states containing a $P$-wave baryon are considerable as well. The state $\Lambda_{c1}|J^P=3/2^-,2\rangle_{\rho}$ may be a broad state with a width of $\Gamma_{\mathrm{Total}}\sim293$ MeV, and mainly decays into $\Sigma_c^*\pi$ and $\Sigma_c\pi$. While, if its mass lie above the threshold of $\Sigma_c\rho$, then the strong decays will be governed by the $\Sigma_c\rho$ channel. As for $\Lambda_{c1}|J^P=5/2^-,2\rangle_{\rho}$, it may be a relatively narrow state with a total width of $\Gamma_{\mathrm{Total}}\sim52$ MeV, and mainly decays into $\Sigma_c^*\pi$. So, this state is most likely to be observed in the $\Lambda_c\pi\pi$ final state via the decay chain $\Lambda_{c1}|J^P=5/2^-,2\rangle_{\rho}\rightarrow \Sigma_c^*\pi\rightarrow \Lambda_c\pi\pi$.

\section*{Acknowledgements }

This work is supported by the National Natural Science Foundation of China under Grants No.12005013, No.12175065, No.12235018 and No.11947048.


	\end{document}